\documentclass{IEEEojcsys}

\usepackage[colorlinks,urlcolor=blue,linkcolor=blue,citecolor=blue]{hyperref}

\usepackage{color,array}

\usepackage{graphicx}

\usepackage{amsmath,amsfonts}
\usepackage{algorithm}
\usepackage{array}
\usepackage{textcomp}
\usepackage{stfloats}
\usepackage{url}
\usepackage{verbatim}
\usepackage{graphicx}
\usepackage{cite}
\usepackage{bm}
\usepackage{mathtools}
\usepackage{xcolor}
\usepackage{upgreek}
\usepackage{tikz}
\usetikzlibrary{arrows.meta, angles, quotes, babel, calc, shapes.misc}
\usepackage{pgfplots}
\usepackage{calrsfs}
\usepackage{amssymb}
\usepackage{subcaption}
\usepackage{csvsimple}
\usepackage{pgfplotstable}
\usepackage{booktabs}
\usepackage{makecell}
\usepackage{cellspace}
\usepackage{multicol}
\usepackage{comment}
\usepackage{threeparttable}
\usepackage{algpseudocode}
\usepackage{nccmath}
\usepackage[export]{adjustbox}
\usepackage{transparent}
\usepackage{scalerel}
\let\labelindent\relax
\usepackage{enumitem}
\usepackage{amsthm}
\usepackage[capitalize]{cleveref}

\DeclareMathAlphabet{\pazocal}{OMS}{zplm}{m}{n}
\DeclareMathOperator*{\argmax}{argmax}
\DeclareMathOperator*{\argmin}{argmin}
\DeclareMathOperator*{\tr}{tr}

\DeclarePairedDelimiter{\mindeg}\lfloor\rfloor
\DeclarePairedDelimiter{\maxdeg}\lceil\rceil

\newcommand{\appropto}{\mathrel{\vcenter{
  \offinterlineskip\halign{\hfil$##$\cr
    \propto\cr\noalign{\kern2pt}\sim\cr\noalign{\kern-2pt}}}}}

\newcommand\given[1][]{\:#1\vert\:}

\newtheoremstyle{mytheoremstyle} % name
    {1pt}                    % Space above
    {1pt}                    % Space below
    {}                   % Body font
    {1em}                           % Indent amount
    {\itshape}                   % Theorem head font
    {:}                          % Punctuation after theorem head
    {.5em}                       % Space after theorem head
    {}  % Theorem head spec (can be left empty, meaning ‘normal’)

\theoremstyle{mytheoremstyle}

\newtheorem{definition}{Definition}
\newtheorem{assumption}{Assumption}
\newtheorem{proposition}{Proposition}
\newtheorem{theorem}{Theorem}
\newtheorem{lemma}{Lemma}

\tikzset{cross/.style={cross out, draw=black, thick, minimum size=2*(#1-\pgflinewidth), inner sep=0pt, outer sep=0pt},
cross/.default={3pt}}

\newenvironment{prooof}{\noindent\hspace{1em}{\itshape Proof: }}{\hfill$\square$}

\newcommand{\norm}[1]{\left\lVert#1\right\rVert}

\jvol{00}
\jnum{XX}
\paper{1234567}
\pubyear{2024}
\receiveddate{XX XX 2024}
\accepteddate{XX XX 2024}
\publisheddate{XX XX 2024}
\currentdate{XX XX 2024}
\doiinfo{OJCSYS.2024.Doi Number}

\begin{document}

\sptitle{Article Category XXX}

\title{Stable Inverse Reinforcement Learning: Policies from Control Lyapunov Landscapes} 

% \editor{This paper was recommended by Associate Editor F. A. Author.}

\author{Samuel Tesfazgi\affilmark{1} (Student Member, IEEE)}

\author{{L}eonhard Sprandl\affilmark{1}}

\author{Armin Lederer\affilmark{2}  (Member, IEEE)}

\author{Sandra~Hirche\affilmark{1} (Fellow, IEEE)}

\affil{Chair of Information-oriented Control (ITR), Technical University of Munich, 80333 Munich, Germany} 
\affil{Learning \& Adaptive Systems Group, ETH Zurich, 8092 Zurich, Switzerland} 

\corresp{CORRESPONDING AUTHOR: Samuel Tesfazgi (e-mail: \href{mailto:samuel.tesfazgi@tum.de}{samuel.tesfazgi@tum.de})}
\authornote{This work was supported by the Horizon 2020 research and innovation program of the European Union under grant agreement no. 871767 of the project ReHyb.}

\markboth{Stable Inverse Reinforcement Learning: Policies from Control Lyapunov Landscapes}{Tesfazgi {\itshape ET AL}.}

\begin{abstract}
Learning from expert demonstrations to flexibly program an autonomous system with complex behaviors or to predict an agent's behavior is a powerful tool, especially in collaborative control settings. A common method to solve this problem is inverse reinforcement learning (IRL), where the observed agent, e.g., a human demonstrator, is assumed to behave according to the optimization of an intrinsic cost function that reflects its intent and informs its control actions. While the framework is expressive, it is also computationally demanding and generally lacks convergence guarantees. We therefore propose a novel, stability-certified IRL approach by reformulating the cost function inference problem to learning control Lyapunov functions (CLF) from demonstrations data. By additionally exploiting closed-form expressions for associated control policies, we are able to efficiently search the space of CLFs by observing the attractor landscape of the induced dynamics. For the construction of the inverse optimal CLFs, we use a Sum of Squares and formulate a convex optimization problem. We present a theoretical analysis of the optimality properties provided by the CLF and evaluate our approach using both simulated and real-world data.
\end{abstract}


\maketitle

\section{INTRODUCTION}

\IEEEPARstart{W}{ith} recent advances in robotic technologies, %and the availability of computational power
autonomous systems are increasingly deployed in new domains. Frequently, these applications include the operation in close proximity and collaboration with human users, such as rehabilitation robotics, manufacturing, and autonomous driving \cite{Mohebbi2020, Hashemi-Petroodi2020, You2019}. As the envisaged tasks and level of collaboration get more involved, %classical adaptive control schemes or compliant impedance controller become insufficient to achieve the desired level of adaptation. Therefore, the need ...
the need to flexibly program expressive human behavior models arises. %, e.g., to generate predictions in a shared control framework \myref{[Predictions]}. 
%In addition to their role as a dynamics predictor, t
%These models are also utilized to flexibly program autonomous systems with complex behaviors, for which the design of a controller is cumbersome and requires expert knowledge. 
The concept of learning a task by observing human experts is referred to as learning from demonstration and imitation learning \cite{Hussein2017, Ravichandar2020} and provides multiple advantages that facilitate effective human-robot collaboration. First, programming complex behavior does not require intimate knowledge regarding the task %control of the autonomous system 
but instead only demonstrations are necessary. Second, user-specific preferences can be integrated in a natural manner, since they are implicitly encoded through the demonstrations. Lastly, the resulting policies and trajectories are easily interpretable as human behavior is emulated \cite{Raza2012}.

Imitation learning is generally more involved than replicating the observed demonstrations exactly via a state to action mapping, since the inferred control policies should ideally generalize well to previously unseen states and unknown environments.
% Beyond generalization characteristics, it is desireable that the solution is also robust to perturbation and small changes in the environment and that the training time and effort is minimized. Finally, the recovered results preferably consitute a rich and insightful representation of the human behavior. There are two principled classes of methods that attempt to account for these desired properties while solving the imitation learning problem. The first approach is called inverse reinforcement learning (IRL), sometimes also referred to as inverse optimal control (IOC), and is originally proposed as an indirect imitation learning method in \myref{[Abbeel\&Ng]}. 
One way to achieve the desired generalization properties is by means of %\mytodos{Alternatively, introduce the high-level requirements/properties desired from IL methods. Then shortly introduce IRL and DMP in 1 sentence each with basic (or only mention that there are) pros and cons. Representation/Complexity $\xleftarrow{}\xrightarrow{}$ Regularity/Convergence Property. } 
inverse reinforcement learning (IRL) %\textcolor{red}{, sometimes also referred to as inverse optimal control (IOC),} 
which is an indirect imitation learning method \cite{Arora2021}. Here, the 
% oprimality principle is the underlying assumption, which is grounded in neuroscientific observations and models an agent to demonstrate optimal behavior with repect to an intrinsic cost function. 
demonstrating agent is assumed to act optimally with respect to an intrinsic cost function. %, which can be used to retrieve the agent's control policy. 
By exploiting this optimality assumption to model the agent, the imitation learning problem is reformulated to inferring the intrinsic cost function driving the agent's actions from observed data \cite{Abbeel2004}. Since the optimal control policy is determined by this cost function, inferring the cost is sufficient to describe the agent's behavior completely. Employing the optimality principle in human motor control is neuroscientfically well justified \cite{Nelson1983, Flash1985} and algorithmically convenient, since it facilitates the use of principled approaches from optimal control theory \cite{Todorov2004}. 
%Precisely, the process by which this inference problem is solved is referred to as IRL and since the control policy is driven by the cost function, inferring it is sufficient to encoding the human behavior. 
Moreover, the cost function constitutes a concise representation of an agent's behavior and has been shown to generalize well to previously unobserved environments \cite{Russell1998}. 
However, in general IRL suffers from the inherent ill-posedness of the inference problem as many cost functions may explain the data well \cite{Ng2000}. While there exist principled approaches to address the ambiguity, such as margin-based \cite{Abbeel2004, Ratliff2006} or entropy-based optimizations \cite{Ziebart2008, ZiebartBrianD.andBagnellJ.AndrewandDey2010, Aghasadeghi2011}, the solution space of potential cost functions remains large. Additionally, IRL methods need to repeatedly solve an optimal control problem in the inner-loop of each learning iteration to evaluate the current cost function parametrization, which is computationally expensive for the nonlinear, nonconvex case.  
%Additionally, it is computationally expensive for nonlinear IRL methods to continuously solve the nonconvex, forward optimal control problem necessary to evaluate the reproductions under the current cost function parametrization. 
To mitigate this issue, some state-of-the-art IRL methods \cite{FinnGCL16, FinnGAN16, Fu17} only solve the forward problem approximately, thereby trading-off computational efficiency and precision. Finally, convergence guarantees for the inferred control policies are typically only provided for the linear-quadratic case \cite{Priess15} or if stable solutions are designed manually \cite{Ornelas12} instead of performing the inference from data.

An alternative and widely used approach to learning by demonstrations is by means of dynamic movement primitives (DMP) \cite{JanIjspeert2002, Schaal2003}. In the DMP framework, demonstrations are represented as a set of differential equation and the resulting, stable dynamical system is used to encode the observed human behavior through its attractor landscape \cite{Pastor2009}. Due to its grounding in dynamical system theory, strong convergence and robustness guarantees can be achieved, which are favorable during operations with humans. Moreover, the model inference can be performed in a computationally efficient manner.
However, in contrast to IRL approaches, DMP methods do not recover a control policy but only provide a dynamical system from which trajectory plans are generated. Therefore, the convenient stability guarantees are limited to the simulated reproductions and are not straight-forwardly retained by a subsequent tracking controller that is required to realize the planned trajectory. Furthermore, since no optimality principle is applied in this framework, DMP methods typically lack the more generalizing model representation provided by an inferred cost function. While some recent approaches concurrently learn a task-oriented potential function, e.g., a Lyapunov-like function, which is used to stabilize the learned dynamical system \cite{Khansari11, Khansari2014, Umlauft2020}, the stabilization is only performed virtually and the control law is not applicable to a physical plant. \looseness=-1 

% Differently to IRL approaches, DMP methods do not recover a control policy but only provide trajectory reproductions, thereby, requiring a subsequent tracking controller to realize the desired movement. 
% However, due to their grounding in dynamical system theory, strong convergence and robustness guarantees can be achieved, which are favorable to ensure safe deployment, in particular during operation with humans. 
% \mytodos{Mention benefits and limitations of DMPs and contrast them to the previous discussion regarding IRL. How to workin related works here? (has to make sense w.r.t. own contribution)}

Therefore, both the IRL and DMP framework solve the imitation learning task in fundamentally different ways, resulting in complementary advantages. %While the DMP approach provides favorable stability properties and facilitates efficient computation, the retrieved dynamical systems representation of the agent's behavior lacks the descriptive richness of a cost function and the practical utility of a control policy, which are obtained in the IRL case. 
In this work we propose an IRL approach that unifies the distinct, beneficial properties of both frameworks. In particular we reformulate the cost function inference problem to learning a Lyapunov-constrained value function approximation from data by exploiting analogies between optimal control theory and robust stabilization, i.e., inverse optimality \cite{Kalman1964, Freeman1996, Krstic1998}. By continuously evaluating the learned value function using a gradient-based stabilizing control law available in closed-form, we directly search over the space of stable, closed-loop dynamical systems that best replicate the observed demonstrations, thereby conceptually bridging the gap to the DMP paradigm.
Preliminary results have appeared in \cite{Tesfazgi2021}, which presents a framework for the time-discrete case showing convergence of the closed-loop
system trajectories to a neighborhood of the equilibrium for a compact state space using a kernel-based approximation. In contrast, this article expands the proposed framework to the time-continuous case and achieves global asymptotic convergence properties. Additionally, a convex optimization approach using sum-of-squares (SOS) is deployed to construct inverse optimal control Lyapunov functions (CLF) from data.
%While there have been works on SOS-based CLFs \cite{Tan2004}, we demonstrate the construction of inverse optimal CLFs using a convex formulation.
Finally, this work presents a rigorous, theoretical analysis of the optimality gap resulting from the Lyapunov-based approximation and provides an extensive evaluation using both simulated and real-world data in a comparison to a state-of-the-art IRL and DMP algorithm. \looseness=-1

\subsection{Organization and Notation}
The remainder of this paper is structured as follows:
After formulating the considered problem in \Cref{sec2}, the Lyapunov-constrained value function approximation is derived in \Cref{sec3}. % together with the results on the optimality gap due to the approximation. 
Following this, \Cref{sec4} investigates the construction of inverse optimal CLFs using SOS and the convexification of the resulting optimization problem. In \Cref{sec5} and \Cref{sec6} the proposed approach is evaluated using both a simulation %, where ground-truth information is available, 
and a comparison study with real-world, human-generated data. Finally, we present our conclusion  in \Cref{sec6}. \looseness=-1

\textit{Notation:} Lower and upper case bold symbols denote vectors and matrices. $\mathbb{R}$ / $\mathbb{N}$ denote the set of all real/integer numbers respectively, while $\mathbb{R}_+$ / $\mathbb{N}_+$ denote all real/integer positive numbers. $\bm{I}_n$ is the $n\times n$ identity matrix. If a matrix $\bm{A}$ is positive definite, we write $\bm{A}\succ\bm{0}$, and $\bm{A}\succeq\bm{0}$ for positive semidefinitness. $\sqrt{\cdot}$ applied to a matrix means it is applied element-wise. %$\|\cdot\|$ the Euclidean norm, and 
$\pazocal{C}$ denotes the set of continuous functions, while $\pazocal{S}$ denotes the set of sum of squares decomposable polynomial functions. $\mindeg{\cdot}$ and $\maxdeg{\cdot}$ denote the lowest and highest degrees of a polynomial's monomials respectively. When applied to a vector of polynomials, $\mindeg{\cdot}$ indicates the minimum and $\maxdeg{\cdot}$ the maximum over the degrees of all elements. 

\section{Problem Formulation} \label{sec2}

Consider a continuous-time, control-affine system
\begin{equation}
	\dot{\bm{x}} = \bm{f}(\bm{x}) + \bm{g}(\bm{x})\bm{u}, \label{eq:unc_dyn}
\end{equation}
with continuous states $\bm{x} \in \mathbb{R}^n$ and initial condition $\bm{x}_0~\in~\mathbb{R}^n$. The system dynamics are given by ${\bm{f}\colon \mathbb{R}^n \to \mathbb{R}^n}$ and ${\bm{g}\colon \mathbb{R}^n \to \mathbb{R}^{n\times m}}$, where $\bm{f(\cdot)}$ and $\bm{g(\cdot)}$ are smooth and \eqref{eq:unc_dyn} is controllable, and the agent acting on the system is assumed to perform continuous control actions $\bm{u}\in \mathbb{R}^m$. 
%Since the agent performing the task is assumed to be an expert, it is reasonable to expect that trajectories generated by its control policies are bounded. 
The system description \eqref{eq:unc_dyn} is not particular restrictive since the assumed control-affine structure holds for many dynamical system classes, such as Euler-Lagrange systems. The following assumption is made for the considered system:
\begin{assumption}
\label{as:known_dyn}
The functions $\bm{f(\cdot)}$ and $\bm{g(\cdot)}$ are known. % and assumed to be polynomial functions.
\end{assumption}

This assumption is not too restrictive since in many application scenarios, e.g., physical human-robot interaction, an accurate dynamics model of the robotic system can either be obtained a-priori using identification techniques \cite{Schoukens2019} or be imposed by appropriate control schemes \cite{Hogan1984}. %Additionally, a polynomial structure is not too restrictive due to their strong uniform approximation capabilities for continuous functions \cite{Stone1948, Ahmadi23}. 
%Differently, the description human behavior models and the predictions of the agent's control actions $\bm{u}$ is a more challenging problem.

%Following findings from behavioral science regarding human motor function \myref{[Hogan84, Todorov04]}
In the following, we employ the optimality principle to describe the behavior of an expert agent. This means that the agent is assumed to choose control actions~$\bm{u}$ according to an intrinsic cost function with the goal of minimizing the accumulated stage costs over time. Thus, the agent employs the optimal control policy
\begin{subequations}
\label{eq:opt_ctl_problem}
    \begin{align} 
    \label{eq:opt_ctl_policy}
    \bm{\pi}^\ast(\bm{x}) = &\argmin_{\bm{\pi}\colon\pazocal{X}\to \pazocal{U}}~ \int_{t=0}^\infty  l(\bm{x} - \bm{x}^\ast) + \bm{\pi}(\bm{x})^\intercal\bm{R}\bm{\pi}(\bm{x})\,dt \\
    &\mathrm{such~that~} \dot{\bm{x}} = \bm{f}(\bm{x}) + \bm{g}(\bm{x})\bm{\pi}(\bm{x}),
\end{align}
\end{subequations}
where $l\colon \mathbb{R}^n\to \mathbb{R}_{+,0}$ is a positive definite function representing state costs and $\bm{R} \in \mathbb{R}^{m\times m} \succ \bm{0}$ is a positive definite matrix denoting the control costs. Due to the structure of $l(\cdot)$ and $\bm{R}$, it is guaranteed that the costs in \eqref{eq:opt_ctl_policy} are positive everywhere except at the target state $\bm{x}^\ast$, which we denote as a meaningful cost in the following according to \cite{Freeman1996}.
% In the following we will only consider the case of quadratic control costs:
% \begin{equation}
%     \label{eq:quadratic_control_costs}
%     q(\bm{u}) = \frac{1}{2}\bm{u}^{\intercal}\bm{R}\bm{u},
% \end{equation}
% where $\bm{R}$ is a positive definite, symmetric matrix. \mynotes{include the quadratic cost later?}

Since human behavior is not deterministic in general, e.g., due to inattentiveness or motor noise \cite{Winter2009}, the agent's control policy does not correspond exactly to the deterministic optimal policy. These unaccounted variations in the control actions can be modelled as random perturbations \cite{Ziebart2008}. Thus, we consider the agent to perform the perturbed optimal control policy
\begin{equation}
    \label{eq:pert_policy}
    \bm{u}dt = \Tilde{\bm{\pi}}(\bm{x})dt \coloneqq \bm{\pi}^\ast\!(\bm{x})dt + d\bm{\omega},
\end{equation}
where $\bm{\omega}$ is an m-dimensional Brownian motion %(sometimes also referred to as Wiener process) 
with state-dependent covariance $\bm{\Sigma}(\bm{x}) \in \mathbb{R}^{m\times m}$. %, which is used to describe the variations. 
Given the perturbed policy \eqref{eq:pert_policy}, the closed-loop dynamics 
% \begin{equation}
%     \label{eq:cl_dyn}
%     d\bm{x} = \bm{f}(\bm{x})dt + \bm{g}(\bm{x})\Big[\bm{\pi}^\ast\!(\bm{x})dt + d\bm{\omega}\Big]
% \end{equation}
are rendered stochastic%. For convenience we rewrite \eqref{eq:cl_dyn} to
\begin{equation}
    \label{eq:cl_dyn_2}
    d\bm{x} = \bm{f}(\bm{x}) + \bm{g}(\bm{x})\bm{\pi}^\ast(\bm{x}) + \bm{\sigma}(\bm{x})d\xi. %\bar{\bm{f}}(\bm{x})dt + \bm{\sigma}(\bm{x})d\xi,
\end{equation}
with %$\bar{\bm{f}}(\bm{x}) \coloneqq \bm{f}(\bm{x}) + \bm{g}(\bm{x})\bm{\pi}^\ast\!(\bm{x})$ and 
$\bm{\sigma}(\bm{x})\coloneqq \bm{g}(\bm{x})\sqrt{\bm{\Sigma}(\bm{x})}$, where $\xi$ indicates n-dimensional standard Brownian motion.
\begin{assumption}
\label{as:known_noise}
The variance $\bm{\Sigma}(\cdot)$ is known.
\end{assumption}
This assumption is in general not particularly restrictive, since $\bm{\Sigma}(\cdot)$ can be identified from data \cite{Hasson2016}. 
While optimal control attempts to solve the problem of finding the optimal policy \eqref{eq:opt_ctl_policy} given the stage costs $l(\cdot)$ and $\bm{R}$, IRL considers the inverse problem. Here, an expert agent demonstrates state-action pairs %$\{(\bm{x_t},\bm{u_t})\}^{T}_{t=0}$, with $T\in\mathbb{N_+}$, 
and the goal is to infer the unknown cost function under which the associated optimal control policy best explains the observations. In order to retrieve the cost function the following assumptions are made for the perturbed policy \eqref{eq:pert_policy} and the closed-loop dynamics \eqref{eq:cl_dyn_2}:

\begin{assumption}
\label{as:unique_target}
The agent performs a goal-directed task, where the goal is defined by a unique target state $\bm{x}^\ast \in \mathbb{R}^n$.
\end{assumption}

\begin{assumption}
\label{as:variance_zero}
The variance in \eqref{eq:cl_dyn_2} vanishes at the target; $\lim_{\bm{x}\to\bm{x}^\ast}\bm{\Sigma}(\bm{x}) = \bm{0}$, such that $\Tilde{\bm{\pi}}(\bm{x}^\ast) = \bm{\pi}^\ast\!(\bm{x}^\ast)$.
\end{assumption}

\cref{as:unique_target} does not pose a strong restriction, because many complex tasks can be achieved by an ordered execution of goal-directed motions \cite{Schaal1999}. Since the target is unique, $\bm{x}^\ast$ coincides with the minimization of the agent's cost function. Moreover, \cref{as:variance_zero} states intuitively that the agent acts more deterministically close to the target state, which is necessary for task completion according to \cref{as:unique_target}.
% \textcolor{cyan}{Intuitively, \cref{as:variance_zero} states that the agent should act more deterministically close to the target state, which is necessary for task completion according to \cref{as:unique_target}. Additionally, \cref{as:unique_target} does not pose a strong restriction, because many complex tasks can be achieved by an ordered execution of goal-directed motions \myref{[Schaal99]}. Since the target is unique, $\bm{x}^\ast$ coincides with the minimization of the agent's cost function. (Switch order of assumption discussion)} 
Therefore, any agent that successfully performs the task has to act on the system such that it asymptotically converges to the desired final state $\bm{x}^\ast$. This has to hold true regardless of the stochasticity due to random perturbations in \eqref{eq:cl_dyn_2}. In order to formalize this property, we introduce the following concept of stability. \looseness=-1
\begin{definition}[\cite{Khalil2002}]
\label{def:stab}
    A system \eqref{eq:unc_dyn} has an asymptotically stable equilibrium $\bm{x}^*$ on the set $\pazocal{X}$ if
    \begin{enumerate}
        \item for all $d \!>\! 0$, there exist  $\delta \!>\! 0$, $t_0\!\geq\! 0$ such that $\norm{\bm{x}_0 \!-\! \bm{x}^*} \!<\! \delta$ implies $\norm{\bm{x}(t) \!-\! \bm{x}^*} \!<\! d $, $\forall t \!\geq\! t_0.$
    \item 
    $\lim_{t \to \infty}\norm{\bm{x}(t) \!-\! \bm{x}^*} \!=\! 0 $ for all $\bm{x}_0\in\pazocal{X}$.
    \end{enumerate}
    If the conditions hold for all states, i.e., $\bm{x}_0\in\mathbb{R}^n$, the equilibrium $\bm{x}^*$ is globally asymptotically stable. 
\end{definition}
% \begin{definition}[\cite{Khasminskii2011}]
% \label{def:stab}
%     A system \eqref{eq:cl_dyn_2} has an asymptotically stable equilibrium $\bm{x}^*$ on the set $\pazocal{X}$ in probability if
%     \begin{enumerate}
%         \item for all $\epsilon\! >\! 0$, $d \!>\! 0$, there exist  $\delta \!>\! 0$, $t_0\!\geq\! 0$ such that $\norm{\bm{x}_0 \!-\! \bm{x}^*} \!<\! \delta$ implies $P\left\{\norm{\bm{x}(t) \!-\! \bm{x}^*} \!<\! d \right\} \geq 1\!-\!\epsilon,$ $\forall t \!\geq\! t_0.$
%     \item 
%     $P\left\{\lim_{t \to \infty}\norm{\bm{x}(t) \!-\! \bm{x}^*} \!=\! 0\right\} \!=\! 1$ for all $\bm{x}_0\in\pazocal{X}$.
%     \end{enumerate}
%     If the conditions hold for all states, i.e., $\bm{x}_0\in\mathbb{R}^n$, the equilibrium $\bm{x}^*$ is globally asymptotically stable in probability. 
% \end{definition}
%Intuitively, \cref{def:stab} is the probabilistic analogue to the classical stability definition in the sense of Lyapunov~\cite{Khalil2002}. %Therefore, we subsequently refer to this property as asymptotic stability. 
% Consequently, using \cref{as:unique_target} and 3, we can express the asymptotic minimization of the objective despite of stochasticity as follows.

Since the task is defined by a unique target state $\bm{x}^*$ in \cref{as:unique_target} at which the policy is deterministic according to \cref{as:variance_zero}, stabilization properties can be imposed for the agent's control policy without loss of generality. %Thereby, additional structure is  

\begin{assumption}
\label{as:stab}
The perturbed policy \eqref{eq:pert_policy} renders the equilibrium $\bm{x}^*$ asymptotically stable. % in probability.% system \eqref{eq:unc_dyn} globally asymptotically stable at the target state $\bm{x}^*$. %=\argmin_{\bm{x}\in\pazocal{X}}V(\bm{x},\tilde{\bm{\pi}}(\bm{x}))$.
\end{assumption}

In general \cref{as:stab} is non-restrictive, since human motion naturally preserves regularity properties \cite{Flash1985}. In the following, we consider the target $\bm{x}^*$ to be the origin without loss of generality. Based on \cref{as:unique_target}, \ref{as:variance_zero} and \ref{as:stab}, we consider the problem of determining the value function
\begin{equation}
    \label{eq:value_function}
    V^*(\bm{x}) \!=\!\! \int_{t_0}^{\infty}\!\! l(\bm{x}(t)) \!+\! \bm{\pi}^\ast\!(\bm{x}(t))^\intercal\bm{R}\bm{\pi}^\ast\!(\bm{x}(t))dt, \enspace \bm{x}_{t_0} = \bm{x} %\mathbb{E}\Bigg[\Bigg],
\end{equation}
which is the minimum cost-to-go from the current state $\bm{x}$ when following the optimal policy $\bm{\pi}^*(\cdot)$. %Here, both $l$ and $\bm{R}$ are unknown and have to be inferred from demonstrations. 
% \textcolor{blue}{Since this data is inherently provided in sampling increments, we can use standard integration methods \myref{[EulerRef]} to approximately write \eqref{eq:cl_dyn_2} to 
% \begin{equation}
%     \label{eq:cl_dyn_ideal}
%     \bm{\dot{x}} = \bar{\bm{f}}(\bm{x}) + \bm{g}(\bm{x})\sqrt{\bm{\Sigma}(\bm{x})}\bm{\varepsilon},
% \end{equation}
% where $\bm{\varepsilon}\sim\pazocal{N}(\bm{0},\bm{I})$ is generated according to a zero-mean Normal distribution.} 
% Here, in contrast to existing approaches, we do not assume access to control measurements, but perform inference only from state observations $\{(\bm{x_t},\bm{\dot{x}_t})\}^{T}_{t=0}$, with $T\in\mathbb{N_+}$. 
The value function $V^*(\cdot)$ is defined though the Hamilton-Jacobi-Bellman (HJB) equation \cite{Fleming2006} 
\begin{equation}
    \label{eq:sHJB}
    \begin{split}
    \nabla_x^\intercal V^*(\bm{x})\bm{f(x)} - \frac{1}{2}\norm{\nabla_x^{\intercal} V^*(\bm{x}) \bm{g}(\bm{x})}_{\bm{R}^{-1}}  + l(\bm{x}) = 0.
    % \nabla_x^\intercal V^*(\bm{x})&\bm{f(x)} - \frac{1}{2}\nabla_x^\intercal V^*(\bm{x}) \bm{g}(\bm{x})\bm{R}^{-1}\bm{g}(\bm{x})^\intercal\nabla_x V^*(\bm{x}) \\ &+ \frac{1}{2}\tr\big(\bm{\sigma}(\bm{x})^\intercal \nabla_{xx}V^*(\bm{x})\bm{\sigma}(\bm{x})\big) + l(\bm{x}) = 0,
    \end{split}
\end{equation}
%where $V^*$ is the solution to the partial differential equation \eqref{eq:sHJB}. 
Moreover, if there exists a continuously differentiable, positive definite solution to \eqref{eq:sHJB}, the optimal feedback control policy $\bm{\pi}^*(\cdot)$ directly follows from $V^*(\cdot)$ as such \cite{Primbs2008}:
\begin{equation}
    \label{eq:opt_feedback}
    \bm{\pi}^*(\bm{x}) = -\bm{R}^{-1}\bm{g}(\bm{x})^\intercal \nabla_x V^*(\bm{x}).
\end{equation}
While solving the HJB \eqref{eq:sHJB} directly is in general not tractable \cite{Primbs2008}, we can exploit the direct dependence of the optimal policy $\bm{\pi}^*(\cdot)$ on the value function $V^*(\cdot)$ in \eqref{eq:opt_feedback} to infer the value function from data. To this end, we use the demonstrations dataset 
\begin{equation}
    \label{eq;data}
    \pazocal{D} = \{\tau_1,\dots,\tau_N \},  \qquad \text{with}\enspace \tau_n=\{\bm{x}_t^{n}\}_{t=0}^T     
\end{equation}
and $T,N \in \mathbb{N}_+$, which encode the agent's preferences. Since the demonstrated state transitions are inherently provided in sampling increments, the dataset is given in a time-discrete form. %Note, in contrast to existing approaches, we do not assume access to control measurements in \eqref{eq;data}, but perform inference only from state observation. 
% we can use standard integration methods \cite{davis2007} to approximately write \eqref{eq:cl_dyn_2} to 
% \begin{equation}
%     \label{eq:cl_dyn_ideal}
%     \bm{\dot{x}} = \bm{f}(\bm{x}) + \bm{g}(\bm{x})\bm{\pi}^\ast(\bm{x}) + \bm{\sigma}(\bm{x})\bm{\varepsilon},
% \end{equation}
% where $\bm{\varepsilon}\sim\pazocal{N}(\bm{0},\bm{I})$ is generated according to a zero-mean Normal distribution.
Together with \cref{as:stab}, we formulate the value function inference as a constrained functional optimization problem
\begin{subequations}
\label{eq:funcopt}
\begin{align}
\label{eq:funcopt1}
    V^* = &\argmax_{V\scaleobj{.9}{\in}\,\pazocal{C}}P\{V\given \pazocal{D} \} \\
    &\mathrm{such~that~} \eqref{eq:opt_feedback} \mathrm{~asymptotically~stabilizes~\eqref{eq:unc_dyn}}.\label{eq:funcopt2}
\end{align}
\end{subequations}

%With the identity \eqref{eq:opt_feedback}, the problem of inferring the stabilizing value function, as conceived in \cref{as:stab}, can now be formulated as a constrained functional optimization problem. 

In \eqref{eq:funcopt1}, we aim to maximize the posterior of $V(\cdot)$ by comparing the state-visitation probabilities of the closed-loop dynamical system under the associated optimal policy and stochastic perturbations with the observed state transitions in the dataset \eqref{eq;data}. By maximizing the posterior $P\{V\given \pazocal{D}\}$ it is possible to include potentially available knowledge through a prior probability distribution. %, where $\tau_n=\{(\bm{x}_t^{n}, \dot{\bm{x}}_t^{n})\}_{t=0}^T$ and $T,N \in \mathbb{N}_+$. 
Thereby, the value function $V^*(\cdot)$, which generates a feedback control $\bm{\pi}^*(\cdot)$ according to \eqref{eq:opt_feedback}, that best explains the demonstrations $\pazocal{D}$ is found. Since additionally \cref{as:stab} must hold true, the posterior maximization becomes a constrained functional optimization problem, with \eqref{eq:funcopt2} guaranteeing that the inferred optimal control policy ${\bm{\pi}^*(\cdot)}$ stabilizes the system.

\section{Lyapunov-based stability-certified inverse reinforcement learning} \label{sec3}
In order to infer the value function $V^\ast(\cdot)$ from data, while also considering the stability constraint on the corresponding optimal policy ${\bm{\pi}}^*(\cdot)$, we exploit the inverse optimal relationship between value functions and CLFs in the following. This allows us to reformulate constraint \eqref{eq:funcopt2} as a Lyapunov constraint on the optimal value function in \cref{subsec:Lyap-Value}. By further considering the optimal value function as a control Lyapunov function and exploiting closed-form expressions for stabilizing control policies, we approximate the posterior maximization \eqref{eq:funcopt1} in a closed-form in \cref{subsec:Likelihood}. Finally, the resulting optimality gap due to the CLF-based approximation of the value function is analyzed theoretically in \cref{subsec:OptimailityGap}. 

\subsection{Lyapunov-Constrained Optimization} \label{subsec:Lyap-Value}
Stability as introduced in \cref{def:stab} and used in \cref{as:stab} is an asymptotic convergence property, which can be analyzed with tools from dynamical system theory. A practical method to ascertain the convergence property of a system, without solving the underlying dynamics equations, is by means of Lyapunov stability theory \cite{Khasminskii2011}. One of the major strengths of Lyapunov stability theory is the existence of converse theorems, i.e., under the assumption of stability, a Lyapunov function is guaranteed to exist \cite{Ahmadi2011}. We exploit this together with another property of Lyapunov functions, which stems from the link between stability and optimality. Namely, the so called inverse optimality property, which states that every Lyapunov function is also a value function for a meaningful optimal stabilization problem. We employ both of these properties to formulate the original problem \eqref{eq:funcopt} as a Lyapunov-constrained optimization problem, which is guaranteed to be feasible. This is shown in the following result.\looseness=-1
\begin{lemma}\label{lem:feas}
The Lyapunov-constrained functional optimization problem
\begin{subequations}
	\label{eq:l1}
\begin{align}
	\label{eq:l1a}
	V^* = &\argmax_{V\scaleobj{.9}{\in}\,\pazocal{C}} &&P\{ V\given  \pazocal{D} \}, \\[3pt]
	\label{eq:l1b}
	& \text{ s.t.}&&    \phantom{\pazocal{L}}V(\bm{x}) \,\; > 0,\enspace 	\forall \bm{x}\in\pazocal{X}\setminus \{\bm{x}^*\}, \\ 
	\label{eq:l1c}
	& &&    \phantom{\pazocal{L}}V(\bm{x}^*) = 0,\enspace \\ 
	\label{eq:l1d}
	& &&  \pazocal{L} V(\bm{x}) \,\; < 0,\enspace	\forall \bm{x}\in\pazocal{X}\setminus \{\bm{x}^*\},
\end{align}
\end{subequations}
is feasible, where 
\begin{equation}
\label{eq:infin_generator}
\begin{split}
    \pazocal{L} V(\bm{x}) = \nabla_{x}V(\bm{x})&^\intercal \big(\bm{f}(\bm{x}) + \bm{g}(\bm{x})\bm{\pi}^\ast(\bm{x})\big) %\\ &+ \frac{1}{2}\tr\big(\bm{\sigma}(\bm{x})^\intercal \nabla_{xx}V(\bm{x})\,\bm{\sigma}(\bm{x})\big)
\end{split}
\end{equation}
is the lie derivative $\pazocal{L} V$ of system \eqref{eq:unc_dyn} under policy $\bm{\pi}^*(\cdot)$. %infinitesimal generator $\pazocal{L} V$  of the stochastic process \eqref{eq:cl_dyn_ideal}.
\end{lemma}
\begin{prooof} 
According to \cref{as:stab} the optimal policy ${\bm{\pi}}^*(\cdot)$ asymptotically stabilizes system \eqref{eq:unc_dyn} in the sense of \cref{def:stab}. Therefore, the converse Lyapunov theorem guarantees the existence of a Lyapunov function satisfying the constraints \eqref{eq:l1b} - \eqref{eq:l1d} \cite{Rajpurohit2016}. %Furthermore, the inverse optimality property establishes that every Lyapunov function also resembles an optimal value function for some meaningful cost defined by $l$ and $R$ in \eqref{eq:value_function} \cite{Freeman1996}, which makes the Lyapunov function a valid solution of \eqref{eq:l1}.
\end{prooof}

Note that the introduction of Lyapunov stability constraints \eqref{eq:l1b} - \eqref{eq:l1d} does not pose a restriction to the solution space, since the inverse optimality property establishes that every CLF also resembles an optimal value function for some meaningful cost defined by $l(\cdot)$ and $\bm{R}$ in \eqref{eq:value_function} \cite{Freeman1996}. Thus, due to the converse Lyapunov theorem and the inverse optimality property, limiting the search to the space of Lyapunov functions still allow to express different preferences of an agent with a stabilizing optimal policy as assumed in \eqref{eq:funcopt2}.

\subsection{Closed-Form Likelihood Expression} \label{subsec:Likelihood}
In the previous section a feasible expression for the stability constraint is derived, which demonstrated the Lyapunov function's principle capability to constitute a solution to the considered functional optimization problem \eqref{eq:funcopt}. In this section we deal with maximizing the posterior \eqref{eq:funcopt1}, thereby ensuring that the found Lyapunov function encodes the observed preferences of the demonstrator.

We follow a Bayesian approach, which directly leads to the proportional relationship
\begin{align}
    P\{V\given  \pazocal{D}\}\propto P\{\tau_1,\ldots,\tau_N\given V  \} P\{V\}, \nonumber
\end{align}
where $P\{V\}$ denotes the prior probability distribution over value functions $V(\cdot)$, which is a design choice. Since the $N$ trajectories $\tau_1,\ldots,\tau_N$ are generated independently using the perturbed optimal policy $\tilde{\bm{\pi}}(\cdot)$, they are conditionally independent given the optimal policy $\bm{\pi}^\ast(\cdot)$. Due to \eqref{eq:opt_feedback}, the associated optimal policy is strictly dependent on $V(\cdot)$, which implies 
\begin{align}
    P\{\tau_1,\ldots,\tau_N\given V  \} = P\{\tau_1\given V \}\cdots P\{\tau_N\given V  \}. \nonumber
\end{align}
Similarly, observed states $\bm{x}_t^n$ along a trajectory $\tau_n$ are conditionally Markovian given the optimal policy $\bm{\pi}^\ast(\cdot)$. Thus, it follows by the same argument as before that %such %that we have for the same argument as before that
\begin{align}
    P\{\tau_n\given V  \} = P\{\bm{x}_0^n\given V\} \prod_{t=1}^T
    P\{{\bm{x}}_t^n\given V, \bm{x}_{t-1}^n \}, \nonumber %P\{\dot{\bm{x}}_t^n\given V, \bm{x}_{t}^n \}, \nonumber
\end{align}
where $P\{\bm{x}_0\given V\}=P\{\bm{x}_0\}$ is the prior initial state distribution. As this prior is independent of $V(\cdot)$, we have
\begin{align}
    P\{\tau_n\given V  \} \propto  \prod_{t=1}^T
    P\{{\bm{x}}_t^n\given V, \bm{x}_{t-1}^n \}. \nonumber %P\{\dot{\bm{x}}_t^n\given V, \bm{x}_{t}^n \}. \nonumber
\end{align}
%Because the agent behaves optimally, it follows that for each of these probabilities, the next observation $\dot{\bm{x}}_{t}^n$ generated by the policy $\bm{\tilde{\pi}}$ applied to the state $\bm{x}_{t}^n$ can be determined using~\eqref{eq:cl_dyn_ideal}. 
%However, in practice we do not consider the stochastic dynamics \eqref{eq:cl_dyn_2} in terms of differentials, but instead obtain measurements with sampling time $\Delta t$, thereby yielding:
% \begin{equation}
%     \label{eq:increment}
%     \Delta\bm{x} = \bar{\bm{f}}(\bm{x})\Delta t + \bm{\sigma}(\bm{x})\Delta\xi. 
% \end{equation}
%Since $\xi$ is generated by standard Brownian motion it follows \mynotes{?from Defintion X?} that $\Delta\xi\sim \pazocal{N}(\bm{0},\Delta t\bm{I}_n)$. Thus, when considering the increment \eqref{eq:increment}, a closed-form formulation for the probabilities can be obtained, resulting in the following expression:
%Since the perturbation $\bm{\varepsilon}$ to the closed-loop dynamics are normally distributed, a closed-form expression for the probabilities can be obtained, resulting in the following expression: 
Since the observed state trajectories are increments sampled from the stochastic process \eqref{eq:cl_dyn_2} with n-dimensional standard Brownian motion $\xi$, it follows that each discrete state-transition probability $P\{{\bm{x}}_t^n\given V, \bm{x}_{t-1}^n \}$ is normally distributed and, using standard integration methods, a closed-form expression for the probabilities can be obtained.
%Since this data is inherently provided in sampling increments, we can use standard integration methods \myref{[EulerRef]} to approximately write
% \begin{equation}
% \begin{split}
\begin{align*}
    & P\{{\bm{x}}_t^n\given V,\bm{x}_{t-1}^n \} = \\  %P\{\dot{\bm{x}}_t^n\given V,\bm{x}_{t}^n \} \!\!=\!
    & \pazocal{N}\big({\bm{x}}_t^n\given \bm{x}_{t-1}^n \!+\! \scaleobj{.8}{\int_{0}^{\Delta t}}\bm{f}(\bm{x}_{t-1}^n) + \bm{g}(\bm{x}_{t-1}^n)\bm{\pi}^\ast(\bm{x}_{t-1}^n)\,d\tau, \\ %\bar{\bm{f}}(\bm{x}_t^n),
   & \hspace{4cm}\bm{g}(\bm{x}_{t-1}^n)^\intercal\bm{\Sigma}(\bm{x}_{t-1}^n)\bm{g}(\bm{x}_{t-1}^n)\big)\nonumber,
\end{align*}
% \end{split}
% \end{equation}}
where $\Delta t$ is the sampling time. By %considering now the trajectories $\tau_n=\{(\bm{x}_t^{n}, \dot{{\bm{x}}}_t^{n})\}_{t=0}^T$ and 
combining the above derived equalities, we obtain the log-likelihood
\begin{align}
\label{eq:exact_loglik}
    \!\log(P\{V\!\given \pazocal{D}\})\propto \log(P\{V\})-\!\sum_{n\!=\!1}^N\sum_{t\!=\!1}^T (\bm{e}_t^n)^\intercal \bm{\Gamma}^{n}_{t}\bm{e}_t^n,\! 
\end{align}
where \allowdisplaybreaks
\begin{align}
    \label{eq:error}
    \bm{e}_t^n& \!=\! \bm{x}_{t}^n \!-\! \Big[ \bm{x}_{t-1}^n \!+\! \scaleobj{.8}{\int_{0}^{\Delta t}}\!\!\bm{f}(\bm{x}_{t-1}^n) \!+\! \bm{g}(\bm{x}_{t-1}^n)\bm{\pi}^\ast(\bm{x}_{t-1}^n)\,d\tau \Big] \\ % \dot{\bm{x}}_t^n-[\bm{f}(\bm{x}_{t}^n)+\bm{g}(\bm{x}_{t}^n)\bm{\pi}^*(\bm{x}_{t}^n)]  \\
    \bm{\Gamma}_t^n&=\left( \bm{g}(\bm{x}_{t-1}^n)^\intercal \,\bm{\Sigma}(\bm{x}_{t-1}^n)\,\bm{g}(\bm{x}_{t-1}^n)\right)^{-1}.
    \label{eq:Gamma}
\end{align}
The log-likelihood \eqref{eq:exact_loglik} measures how well the perturbed optimal policy $\bm{\tilde{\pi}}(\cdot)$ approximates the observed demonstrations $\pazocal{D}$
of the agent. Thus, the optimal value function $V^*(\cdot)$ is retrieved, as the unperturbed optimal policy $\bm{\pi}^*(\cdot)$ in \eqref{eq:error} directly depends on $V(\cdot)$ due to \eqref{eq:opt_feedback}. Thereby, the loss in \eqref{eq:exact_loglik} is minimized by finding a Lyapunov candidate function $V(\cdot)$ with an associated optimal policy $\bm{\pi}^*(\cdot)$ \eqref{eq:opt_feedback} that generates closed-loop dynamics with a state-visitation probability similar to the observed demonstration trajectories $\tau_1,\ldots,\tau_N$. \looseness=-1
%best explains the demonstrations.

However, the likelihood maximization \eqref{eq:exact_loglik} is done with respect to a finite demonstration dataset $\pazocal{D}$ and therefore may only locally reproduce the true, underlying value function. Thus, we denote the resulting estimated value function with $\hat{V}(\cdot)$ in the following. 
In contrast, the desired stabilization property \eqref{eq:funcopt2} of the associated control policy \eqref{eq:opt_feedback} %, formalized in \eqref{eq:l1b} - \eqref{eq:l1d}, 
must hold true for an uncountable, infinite set of states in $\pazocal{X}$ and is not limited to the finite set of observed data points $\pazocal{D}$. Therefore, while the optimal control policy $\bm{\pi}^*(\cdot)$ induced by the true, unknown value function $V^*(\cdot)$ renders the system asymptotically stable due to \Cref{as:stab}, this may generally not be true when applying control law \eqref{eq:opt_feedback} with an estimated value function $\hat{V}(\cdot)$. %In particular, this is not only due to the finite amount of available observations, but also because the observed state transitions ... are generated by the perturbed optimal policy $\bm{\tilde{\pi}}$. 
To overcome this, we make use of the fact that the estimated $\hat{V}(\cdot)$ is a Lyapunov function by design due to the constraints \eqref{eq:l1b} - \eqref{eq:l1d}, from which two advantages follow. First, the existence of a CLF for a control-affine system, such as \eqref{eq:unc_dyn}, implies the existence of an asymptotically stabilizing control law \cite{Artstein1983}, which we will denote with  $\hat{\bm{\pi}}(\bm{x})$ in the following. Secondly, it is possible to calculate such a stabilizing control law $\hat{\bm{\pi}}(\cdot)$ explicitly and in closed-form if the system dynamics and CLF are known \cite{Sontag1989}. 
In \cite{Freeman96cdc} a slight variation of the original formulation (Sontag's formula \cite{Sontag1989}) is introduced
\begin{equation}
    \hat{\bm{\pi}}(\bm{x}) \!=\! 
    \begin{cases}
        -\lambda(\bm{x})\bm{R}^{-1}\bm{g}(\bm{x})^\intercal  \nabla_x V(\bm{x}), & \quad \bm{g}^\intercal  \nabla_x V \neq 0\\
        0,              & \quad \bm{g}^\intercal  \nabla_x V = 0
    \end{cases}
    \label{eq:sontag}
\end{equation}
where $\lambda(\bm{x})\colon \mathbb{R}^n\to \mathbb{R}_{+,0}$ is a state-dependent, positive and continuous scalar factor. While in this work we propose to consider $\lambda(\cdot)$ as a functional design parameter, thereby softening the coupling between optimal and stabilizing policy in favor of gaining additional freedom for the control design, there are also analytic constructions of the scaling factor \cite{Freeman96cdc}
\begin{equation}
    \lambda(x) = \cfrac{\nabla_{x}^{\intercal}V\bm{f} \!+\! \sqrt{(\nabla_{x}^{\intercal}V\bm{f})^2 + l(\bm{x})\big( \nabla_x^{\intercal} V \bm{g}\bm{g}^\intercal\nabla_x V\big) } }{\nabla_x^{\intercal} V \bm{g}\bm{g}^\intercal\nabla_x V},
    \label{eq:lambda_explicit}
\end{equation}
where we omit the dependency on $\bm{x}$ for $\bm{f}$, $\bm{g}$ and $V$ for improved readability. It is shown in \cite{Primbs2008} that the control formulation \eqref{eq:sontag} together with \eqref{eq:lambda_explicit} produces the optimal policy $\bm{\pi}^*(\cdot)$ exactly, if the CLF $V(\cdot)$ and value function $V^*(\cdot)$ have the same shape level sets, i.e., the only difference between their gradients is a state dependent, scalar proportional factor
\begin{equation}
    \label{eq:same_level}
    \nabla_x V^* = \lambda(\bm{x}) \nabla_x V,
\end{equation}
which can be seen by substituting \eqref{eq:same_level} into the HJB \eqref{eq:sHJB} and solving for $\lambda(\cdot)$. Intuitively, the stabilizing policy $\hat{\bm{\pi}}(\cdot)$ in \eqref{eq:sontag} is similar to the optimal policy $\bm{\pi}^*(\cdot)$ \eqref{eq:opt_feedback} in the sense that both follow the directional information provided by the gradient of a potential function, e.g., either the the CLF $V(\cdot)$ or value function $V^*(\cdot)$. Therefore, given a CLF $V(\cdot)$, employing control law \eqref{eq:sontag} stabilizes system \eqref{eq:unc_dyn}, whilst an appropriate choice for the point-wise scaling $\lambda(\cdot)$ along the gradient descent direction of $V(\cdot)$ may additionally solve the HJB equation. Hence, instead of employing the relationship between the value function $V^*(\cdot)$ and optimal policy $\bm{\pi}^*(\cdot)$, we propose to utilize the stabilizing control law $\hat{\bm{\pi}}(\cdot)$, since the estimated $\hat{V}(\cdot)$ is already constrained to be a CLF.

Analogously to before, the  CLF $\hat{V}(\cdot)$ is fitted by evaluating the incurred loss under the closed-form control law $\hat{\bm{\pi}}(\cdot)$ %This immediately leads to the approximation \looseness=-1
\begin{align}
\label{eq:approx_loglik}
    \!\log(P\{\hat{V}\!\given \pazocal{D}\})\appropto \log(P\{\hat{V}\})-\!\sum_{n\!=\!1}^N\sum_{t\!=\!1}^T (\hat{\bm{e}}_t^n)^\intercal\bm{\Gamma}^{n}_t\hat{\bm{e}}_t^n,\!
\end{align}
where 
\begin{align}
    \bm{e}_t^n& \!=\! \bm{x}_{t}^n \!-\! \Big[ \bm{x}_{t-1}^n \!+\! \scaleobj{.8}{\int_{0}^{\Delta t}}\!\!\bm{f}(\bm{x}_{t-1}^n) \!+\! \bm{g}(\bm{x}_{t-1}^n)\hat{\bm{\pi}}(\bm{x}_{t-1}^n)\,d\tau \Big] %\hat{\bm{e}}_t^n&=\dot{\bm{x}}_t^n-[\bm{f}(\bm{x}_{t}^n)+\bm{g}(\bm{x}_{t}^n)\hat{\bm{\pi}}(\bm{x}_{t}^n)].
    \label{eq:error_hat}
\end{align}
This approximate log-likelihood allows to infer ${V^*}(\cdot)$ indirectly by fitting a policy, merely substituting the optimal policy $\bm{\pi}^*(\cdot)$ by the stabilizing policy $\hat{\bm{\pi}}(\cdot)$. Thereby the proposed optimization \eqref{eq:approx_loglik} still yields an estimate $\hat{V}(\cdot)$, which reflects the agent's preferences, since the associated stabilizing policy $\hat{\bm{\pi}}(\cdot)$ best possibly represents the observed demonstrations, even though $\hat{V}(\cdot)$ is not an optimal value function in general. %, the formulation in \eqref{eq:approx_loglik} still allows to encode %arbitrary
%agent preferences in principle. % and can in some cases induce the optimal policy $\bm{\pi}^*$ exactly.

\subsection{Inverse Optimal Value Function Approximation} \label{subsec:OptimailityGap}
%So far an approximate IRL approach is presented that substitutes the value function $V^*$ and optimal policy $\bm{\pi}^*$ by optimizing over a CLF $V$ and a stabilizing control law $\hat{\bm{\pi}}$. 
In this section we perform a theoretical analysis of the optimality gap resulting due to the approximation of $V^*(\cdot)$ using the CLF $\hat{V}(\cdot)$. %\textcolor{red}{and demonstrate under which conditions the obtained approximation is exact.} 
Recall that the observed agent is modelled using the optimality principle, i.e., there exists an intrinsic cost function \eqref{eq:opt_ctl_policy} that the agent minimizes by employing the optimal policy $\bm{\pi}^*(\cdot)$. The value function $V^*(\cdot)$, that we attempt to infer in \eqref{eq:funcopt1}, represents an optimal aggregate of the costs experienced by the agent starting in any current state to the goal state. Thus, substituting the value function $V^*(\cdot)$ with the CLF-based approximation $\hat{V}(\cdot)$ also influences the underlying cost function that is being minimized by the associated policy $\hat{\bm{\pi}}(\cdot)$ and thereby impacts the representation of the modelled agent. %the \textcolor{red}{inferred cost function representation} of the modelled agent, which we investigate in the following.}

To theoretically analyze the implications of employing the CLF $\hat{V}(\cdot)$ and stabilizing policy $\hat{\bm{\pi}}(\cdot)$, we insert the state-dependent scaling factor $\lambda(\cdot)$ according to \eqref{eq:lambda_explicit}. For notational convenience we let
    \begin{equation}
        a(\bm{x}) = \nabla_x^{\intercal} \hat{V}(\bm{x})\bm{f}(\bm{x}), \label{eq:a}
    \end{equation}
    and
    \begin{equation}
        b(\bm{x}) = \norm{\nabla_x^{\intercal} \hat{V}(\bm{x}) \bm{g}(\bm{x})}_{\bm{R}^{-1}}, \label{eq:b} %b(\bm{x}) = \nabla_x^{\intercal} V \bm{g}\bm{R}^{-1}\bm{g}^\intercal\nabla_x V, \label{eq:b}
    \end{equation}   
and introduce the slightly modified scaling function 
    \begin{equation}
        \lambda(\bm{x}) = \frac{a(\bm{x}) + \sqrt{a(\bm{x})^2 + 2b(\bm{x})l(\bm{x}) } }{b(\bm{x})}. \label{eq:lambda_x}
    \end{equation}
When employing the CLF-based approximation to model the agent, the policy $\hat{\bm{\pi}}(\cdot)$ induced by the CLF $\hat{V}(\cdot)$ may not constitute an optimal minimizer for the cost function \eqref{eq:opt_ctl_policy} in general. However, the proposed formulation still retains a strong connection to the original optimal control problem \eqref{eq:opt_ctl_problem}, which is shown in the following result.

\begin{theorem} \label{th:optimality}
    Consider a control-affine system as in \eqref{eq:unc_dyn} and a CLF-based value function estimate $\hat{V}(\cdot)$. Then applying control policy $\hat{\bm{\pi}}(\cdot)$ \eqref{eq:sontag} with scaling factor $\lambda(\cdot)$ according to \eqref{eq:lambda_x} solves the modified optimal control problem
    \begin{subequations}
    \label{eq:inv_opt_ctl_problem}
    \begin{align} 
        \hat{\bm{\pi}}(\bm{x}) = &\argmin_{\bm{\pi}\colon\pazocal{X}\to \pazocal{U}}~~ \int_{t=0}^\infty  \hat{l}(\bm{x}) + \hat{r}(\bm{x}, \bm{\pi(\bm{x})})\,dt \\
        &\mathrm{such~that~} \dot{\bm{x}} = \bm{f}(\bm{x}) + \bm{g}(\bm{x})\bm{\pi}(\bm{x}),
    \end{align}
    \end{subequations}
    with the meaningful cost functionals
    \begin{equation} 
        \hat{l}(\bm{x}) = \frac{1}{2}\lambda(\bm{x})b(\bm{x}) - a(\bm{x}), \label{eq:l_hat}
    \end{equation}
    and
    \begin{equation}
        \hat{r}(\bm{x}) = \frac{1}{\lambda(\bm{x})}\bm{\pi}(\bm{x})^\intercal\bm{R}\bm{\pi}(\bm{x}). \label{eq:q_hat}
    \end{equation}
    Further, if $\hat{V}(\cdot)$ and $V^*(\cdot)$ have the same shape level sets \eqref{eq:same_level}, the modified optimal control problem \eqref{eq:inv_opt_ctl_problem} is similar to the original problem \eqref{eq:opt_ctl_problem} in the sense that the relation
    \begin{equation}
        \hat{l}(\bm{x})r(\bm{x}) =  l(\bm{x})\hat{r}(\bm{x}), %\frac{l(\bm{x})}{q(\bm{x})} = \frac{\hat{l}(\bm{x})}{\hat{q}(\bm{x})}, \label{eq:cost_fct_relation}
        \label{eq:cost_fct_relation}
    \end{equation}
    where $r(\bm{x})=\bm{\pi}(\bm{x})^\intercal\bm{R}\bm{\pi}(\bm{x})$ denotes the original control costs in \eqref{eq:opt_ctl_policy}, holds true.
\end{theorem}
\begin{prooof}
    In order to show that $\hat{\bm{\pi}}(\cdot)$ solves the modified optimal control problem \eqref{eq:inv_opt_ctl_problem}, we insert \eqref{eq:l_hat} in the HJB%, which yields
    \begin{equation}
    \begin{split}
        % \nabla_x V^{\intercal}\bm{f}  - \frac{\lambda}{2}\nabla_x V^{\intercal} \bm{g}\bm{R}^{-1}\bm{g}^\intercal\nabla_x V + \hat{l}(\bm{x}) &= 0 \nonumber \\
        \nabla_x^{\intercal} V^*(\bm{x})\bm{f}(\bm{x}) \!-\! \frac{\lambda(\bm{x})}{2}\!&\norm{\nabla_x^{\intercal} V^*(\bm{x}) \bm{g}(\bm{x})}_{\bm{R}^{-1}} \!\! \\
        &\qquad +\! \frac{\lambda(\bm{x})}{2}b(\bm{x}) \!-\! a(\bm{x}) \!=\! 0 \label{eq:mod_HJB}. %\nabla_x &V^{\intercal} \bm{g}\bm{R}^{-1}\bm{g}^\intercal\nabla_x V \\  &+ \frac{1}{2}\lambda(\bm{x})b(\bm{x}) - a(\bm{x}) = 0 \label{eq:mod_HJB}.
    \end{split}        
    \end{equation}
    With \eqref{eq:a} and \eqref{eq:b}, it can be seen that the equality \eqref{eq:mod_HJB} holds, thereby showing that $\hat{\bm{\pi}}(\cdot)$ is the optimal policy for problem \eqref{eq:inv_opt_ctl_problem}. For $\hat{l}(\cdot)$ and $\hat{r}(\cdot)$ to constitute a meaningful cost, it remains to prove their positive definiteness. Since both $b(\cdot)$  and $l(\cdot)$ are positive definite it follows directly that
    \begin{equation}
        \sqrt{a(\bm{x})^2 +2b(\bm{x})l(\bm{x})} \geq a(\bm{x}), \quad\forall x\in\mathbb{R}^n, \label{eq:larger_root}
    \end{equation}
    when taking only the positive square root. This directly implies that $\lambda(\cdot)$ is positive definite according to \eqref{eq:lambda_x}. Together with $\bm{R}\succ\bm{0}$ it follows that $\hat{r}(\cdot)$ in \eqref{eq:q_hat} is positive definite. Now inserting \eqref{eq:lambda_x} into \eqref{eq:l_hat} we get
    \begin{equation}
        \hat{l}(\bm{x}) = \frac{\sqrt{a(\bm{x})^2 + 2b(\bm{x})l(\bm{x})} - a(\bm{x}) }{2},
    \end{equation}
    which is also positive definite due to \eqref{eq:larger_root}. Therefore, the modified optimal control problem \eqref{eq:inv_opt_ctl_problem} has a meaningful cost function and the inverse optimality property holds.

    For the second part, it remains to demonstrate that \eqref{eq:cost_fct_relation} is true. %, which can be written equivalently to:
    % \begin{equation}
    %     \hat{l}(\bm{x})q(\bm{x}) - l(\bm{x})\hat{q}(\bm{x}) = 0.
    % \end{equation}
    By inserting \eqref{eq:l_hat} and \eqref{eq:q_hat} and writing out $r(\cdot)$ we obtain
    \begin{align}
        % \Big( \frac{1}{2} \lambda b - a \Big)\frac{1}{2}\bm{\pi}(\bm{x})^\intercal\bm{R}\bm{\pi}(\bm{x}) - \frac{l}{2\lambda(\bm{x})}\bm{\pi}(\bm{x})^\intercal\bm{R}\bm{\pi}(\bm{x}) &= 0 \nonumber \\
        \Big( \frac{1}{2} \lambda(\bm{x})^2 b(\bm{x}) \!-\! \lambda(\bm{x}) a(\bm{x}) \!-\! l(\bm{x})\Big) \frac{1}{2\lambda(\bm{x})}\bm{\pi}(\bm{x})^\intercal\bm{R}\bm{\pi}(\bm{x}) &= 0, \nonumber
    \end{align}
    which holds true for $\lambda(\cdot)$ satisfying 
    \begin{equation}
        \frac{1}{2} \lambda(\bm{x})^2 b(\bm{x}) - \lambda(\bm{x}) a(\bm{x}) - l(\bm{x}) = 0. \label{eq:lambda_root}
    \end{equation}
    If the CLF $\hat{V}(\cdot)$ and value function $V^*(\cdot)$ have co-linear gradients, we can rewrite the HJB \eqref{eq:sHJB} in terms of $\lambda(\bm{x})\nabla_x \hat{V}(\cdot)$ such that
    %Since the CLF $V$ and value function $V^*$ have co-linear gradients and Hessians, we can rewrite the stochastic HJB \eqref{eq:sHJB} by substituting $\nabla_x V^*$ and $\nabla_{xx}V^*$ with the scaled $\nabla_x V$ and $\nabla_{xx}V$ such that
    \begin{equation}
        \label{eq:inverse_optimal_HJB}
        \begin{split}
        \lambda(\bm{x})\nabla_x^{\intercal} \hat{V}(\bm{x})\bm{f}(\bm{x}) - \frac{1}{2}\lambda(\bm{x})^2&\norm{\nabla_x^{\intercal} \hat{V}(\bm{x}) \bm{g}(\bm{x})}_{\bm{R}^{-1}} \\ &\qquad\qquad+ l(\bm{x}) = 0.
        %\lambda(\bm{x})\Big(\nabla_x &V^{\intercal}\bm{f} + \frac{1}{2}\tr\big(\bm{\sigma}^\intercal \nabla_{xx}V\bm{\sigma}\big) \Big)  \\ &- \frac{\lambda(\bm{x})^2}{2}\nabla_x V^{\intercal} \bm{g}\bm{R}^{-1}\bm{g}^\intercal\nabla_x V + l(\bm{x}) = 0.
        \end{split}
    \end{equation}
    Thus, given the definition of $a(\cdot)$ \eqref{eq:a} and $b(\cdot)$ \eqref{eq:b}, it follows from \eqref{eq:inverse_optimal_HJB} that equality \eqref{eq:lambda_root} holds true, %for the case that the CLF $V$ and value function $V^*$ have the shape of level sets as described in \eqref{eq:same_level}
    which concludes the proof.
\end{prooof}

Intuitively, the similarity \eqref{eq:cost_fct_relation} in \cref{th:optimality} indicates that the proposed approximation via the CLF $\hat{V}(\cdot)$ and stabilizing policy $\hat{\bm{\pi}}(\cdot)$ weights the state and control costs proportionally in the same way as in the original optimal control problem \eqref{eq:opt_ctl_problem}. %Additionally, both the applied policy $\hat{\bm{\pi}}$ and the optimal policy $\bm{\pi}^*$ follow the gradient direction of $V$ and $V^*$ respectively, thereby inducing co-linear shape of the two potential functions. 
%Therefore, the inferred $V$ and $\hat{\bm{\pi}}$ can be expected to qualitatively resemble the optimal $V^*$ and $\bm{\pi}^*$ and induce similar behavior. 
Another way to interpret this correspondence is that the modified optimal control problem \eqref{eq:inv_opt_ctl_problem} modulates the stage costs of the original optimal control problem \eqref{eq:opt_ctl_problem} using $\lambda(\cdot)$:
\begin{equation}
    \hat{\bm{\pi}}(\bm{x}) = \argmin_{\bm{\pi}\colon\pazocal{X}\to \pazocal{U}}~~ \int_{t=0}^\infty  \frac{1}{\lambda(\bm{x})}\big(l(\bm{x}) + r(\bm{x})\big)\,dt, \nonumber
\end{equation}
which can be seen by multiplying $\hat{l}(\cdot)$ \eqref{eq:l_hat} and $\hat{r}(\cdot)$ \eqref{eq:q_hat} with $\lambda(\cdot)$ and solving the equations using \eqref{eq:lambda_x}.

\section{Sum of Squared based Lyapunov formulation}\label{sec4}
While the formulation in \eqref{eq:l1} together with the approximate log-likelihood \eqref{eq:approx_loglik} provides an appealing framework for the inference of $V^*(\cdot)$, it requires solving a functional optimization problem. Therefore, expressive function approximators are needed. %directly solving the functional optimization problem is intractable in practice. 
Thus, in this section we employ the sum of squares (SOS) technique %\cite{Parillo00, Papachristodoulou2002} 
and transform \eqref{eq:l1} into a constrained optimization over the space of polynomials. %\textcolor{cyan}{Since the resulting formulation is intractable for conventional solvers, we further transform the problem into semi-definite programs (SDP) which can be solved efficiently.} 
%\cref{subsec:sos-decomp} provides a short introduction to SOS programming. 
In \cref{subsec:sos-optimization} we rewrite the functional optimization \eqref{eq:l1} into an SOS program and introduce Lyapunov constraints in \cref{subsec:sos-lya-constr}. Finally, we propose a convexification scheme in \cref{subsec:sos-convexify} to create tractable SDPs for the Lyapunov-constrained optimization. %from the result of the previous section. 

\subsection{Formulation as SOS Program} \label{subsec:sos-optimization}
The SOS technique is an efficient approach to test whether a given polynomial is non-negative by checking if it can be expressed as a sum of squares. Since the Lyapunov constraints \eqref{eq:l1b} - \eqref{eq:l1d} are also positive-definiteness conditions, they can conveniently be expressed as SOS constraints \cite{Parillo00}. %The previous subsection provides a convenient approach to express the positive definiteness constraints of the Lyapunov function in the form of SOS polynomials. 
To this end, we parameterize $\hat{V}(\cdot)$ as a multivariate polynomial consisting of monomial vector $\bm{z}(\bm{x})$ and real scalar coefficients $\bm{c}$
%When using the SOS formalism, the search for $V^*(\bm{x})$ reduces to a search over the coefficients $\bm{c}$ that parameterize a multivariate polynomial candidate $V$:
\begin{equation}
    \label{eq:V_poly}
    \hat{V}(\bm{c},\bm{x})= \bm{c}^\intercal \bm{z}(\bm{x}),
\end{equation}
% where \vspace{-0.1cm}
% \begin{align}
%     \bm{c} & := [c_1,\dots,c_j]^\intercal \text{,} \\
%     \bm{z}(\bm{x}) & := [z_1(\bm{x}), \dots, z_j(\bm{x})]^\intercal
% \end{align}}
which is not particularly restrictive %due to \cref{as:known_dyn} and 
since polynomials are known to be sufficiently expressive to approximate a large family of functions \cite{Ahmadi23}.
%Here, $\bm{z}(\bm{x})$ denotes a vector of monomials with degree $d_m$ where $d_m \geq 2$. 
Due to \eqref{eq:V_poly}, the stabilizing control law $\hat{\bm{\pi}}(\cdot)$ becomes
\begin{equation}
    \hat{\bm{\pi}}(\bm{c},\bm{x}) \!=\! 
    \begin{cases}
        -\lambda(\bm{x})\bm{R}^{-1}\bm{g}^\intercal  \nabla_{x}\bm{z}(\bm{x})^\intercal\bm{c}, & \enspace \bm{g}^\intercal  \nabla_x \bm{z} \!\neq\! 0\\
        0,              & \enspace \bm{g}^\intercal  \nabla_x \bm{z} \!=\! 0,
    \end{cases}
 \label{eq:c_sontag}
\end{equation}
which has a linear dependence on $\bm{c}$. Thus, the empirical error \eqref{eq:error_hat} admits a representation that is linear in $\bm{c}$ as well
\begin{equation}
    \label{eq:param_error}
    \begin{split}
    \hat{\bm{e}}_{t+1}^n(\bm{c}) = \bm{x}&_{t+1}^n \!-\! \bm{x}_{t}^n \\ &\!+\!  \scaleobj{.8}{\int_{0}^{\Delta t}}\!\!\bm{f}(\bm{x}_{t}^n) \!+\! \lambda(\bm{x}_{t}^n)\!\norm{\bm{g}(\bm{x}_{t}^n)}_{\bm{R}^{-1}}\!\! \nabla_{x}^\intercal\bm{z}(\bm{x}_{t}^n)\bm{c}\,d\tau \nonumber
    \end{split}
    %= \dot{\bm{x}}_t^n - \big[\bm{f}(\bm{x}_t^n) - \lambda(\bm{x})\bm{g}(\bm{x}_t^n)\bm{g}(\bm{x}_t^n)^\intercal\nabla_{x}\bm{z}(\bm{x}_t^n)^\intercal\bm{c} \big].
\end{equation}
Moreover, since the posterior optimization \eqref{eq:approx_loglik} reduces to a search over the coefficients $\bm{c}$, the prior $P(\hat{V})$ in \eqref{eq:approx_loglik} can be defined in terms of $\bm{c}$ as well. We choose a multivariate, zero-mean normal distribution ${\bm{c}\sim \pazocal{N}(\bm{0},\bm{\Sigma}_c)}$, which yields a closed-form expression for the parameterized log-likelihood
\begin{equation}
    \label{eq:param_loglik}
        \!\log(P\{\hat{V}\!\given \pazocal{D}\})\!\appropto\! -\frac{1}{2}\bm{c}^\intercal\bm{\Sigma}_c^{-1}\bm{c} -\sum_{n\!=\!1}^N\sum_{t\!=\!1}^T \hat{\bm{e}}_t^n(\bm{c})^\intercal\bm{\Gamma}^{n}_t\hat{\bm{e}}_t^n(\bm{c}).\!
    % \begin{split}
    % \!\log(P\{V\!\given \pazocal{D}\})\!\appropto\! &-\frac{1}{2}\bm{c}^\intercal\bm{\Sigma}_c^{-1}\bm{c} \\ &\quad-\sum_{n\!=\!1}^N\sum_{t\!=\!1}^T \hat{\bm{e}}_t^n(\bm{c})^\intercal\bm{\Gamma}^{n}_t\hat{\bm{e}}_t^n(\bm{c}).\!
    % \end{split}
\end{equation}
The parameterized log-likelihood \eqref{eq:param_loglik} contains terms that are quadratic in $\bm{c}$ and can therefore not be used directly in a convex SDP problem. %, as here the objective is required to be linear in the decision variables \cite{Boyd04}. 
%However, in the following result we demonstrate that the maximization of the likelihood \eqref{eq:param_loglik} can be reformulated equivalently as an SDP problem.
However, using the epigraph formulation, \eqref{eq:param_loglik} is reformulated to be linear in the decision variables.
%However by using an equivalent representation, the maximization of the likelihood \eqref{eq:param_loglik} can be reformulated as an SDP problem, which is demonstrated in the following result.
% Hence, evaluating this error for a data-point $(\bm{x}_t^n, d\bm{x}_t^n)$ yields a measure of similarity between human demonstrations and the closed-loop system reproductions given a parameterization of a candidate $V$ defined by parameter vector $\bm{c}$. Having a closed-form expression of the error $\hat{e}$ \eqref{eq:e_explicit} that is linear in $\bm{c}$ is beneficial, because it allows us to formulate the problem of maximizing the approximate log-likelihood \eqref{eq:approx_loglik} as an SDP problem, which requires a linear dependence on the decision variables as shown in \eqref{eq:generic-sdp}.
\begin{lemma}\label{lem:err_schur}
    Consider a control-affine system \eqref{eq:unc_dyn}, a dataset $\pazocal{D}$ as in \eqref{eq;data}, %${D\!=\!\{\tau_1,\dots,\tau_N\}}$ of $N$ trajectories ${\tau_n=\{(\bm{x}_t^{n}, \dot{\bm{x}}_t^{n})\}_{t=0}^T}$
    a control law $\bm{\hat{\pi}}(\bm{c}, \bm{x})$ according to \eqref{eq:c_sontag} and the scalar variables $q,p\in\mathbb{R}$. Then maximizing the parameterized log-likelihood \eqref{eq:param_loglik} over $\pazocal{D}$ is equivalent to solving the SDP problem
    \begin{subequations}
        \label{eq:sdp-obj}
        \begin{align}
            \label{eq:sdp-obj-a}
            \bm{c^*} = &\argmin_{\bm{c}, q, p} \quad q+p \\
            \label{eq:sdp-obj-b}
            &\text{ such that} \enspace \bm{Q}(q, \bm{c}) \succeq 0,  \\
            & \phantom{\text{ such that}} \enspace \bm{P}(p, \bm{c}) \succeq 0,
        \end{align}
    \end{subequations}
    where
    \begin{align}
        \label{eq:sdp-obj-def-P}
        \bm{P}(p, \bm{c}) := &\begin{bmatrix}
            p & \bm{c}^\intercal \\
            \bm{c} & \bm{\Sigma}_c
        \end{bmatrix} \in \mathbb{R}^{\binom{n+d_m}{d_m} \times \binom{n+d_m}{d_m}}, \\[5pt]
        \label{eq:sdp-obj-def-Q}
        \bm{Q}(q, \bm{c}) := &\begin{bmatrix}
            q & \bm{\varepsilon}(\bm{c}) \\
            \bm{\varepsilon}(\bm{c})^\intercal & \bm{G}^{-1}
        \end{bmatrix} \in \mathbb{R}^{(1+nNt) \times (1+nNt)},
    \end{align}
    with
    \begin{align}
            \label{eq:sdp-obj-def-S}
            \bm{\varepsilon}(\bm{c}) := & \big[(\hat{\bm{e}}_1^1(\bm{c}))^\intercal ... (\hat{\bm{e}}_t^n(\bm{c}))^\intercal \big] \in \mathbb{R}^{1 \times nNt} \\[5pt]
            \label{eq:sdp-obj-def-R}
            \bm{G} := & \text{diag}\big(\bm{\Gamma}^{1}_1, \ldots, \bm{\Gamma}^{n}_t\big) \in \mathbb{R}^{(nNt) \times (nNt)}.
            % \bm{G} := & \begin{bmatrix}
            %     \bm{\Gamma}^{1}_1 & & \\
            %      & \ddots & \\
            %     & & \bm{\Gamma}^{n}_t
            % \end{bmatrix} \in \mathbb{R}^{nNt \times nNt}.
    \end{align}
    % and
    % \begin{equation}
    %     \label{eq:sdp-obj-def-P}
    %     \bm{P}(p, \bm{c}) := \begin{bmatrix}
    %         p & \bm{c}^\intercal \\
    %         \bm{c} & \bm{\Sigma}_c
    %     \end{bmatrix} \in \mathbb{R}^{\binom{n+d_m}{d_m} \times \binom{n+d_m}{d_m}},
    % \end{equation}
\end{lemma}

\begin{prooof}
    We maximize \eqref{eq:param_loglik} %Maximizing the parameterized log-likelihood \eqref{eq:param_loglik} is done 
    by minimizing the regularized empirical error $\bm{\hat{e}}(\bm{c})$ over dataset $\pazocal{D}$ \looseness=-1
    \begin{equation}
        \label{eq:obj-reform-f1}
        \bm{{c}^*} = \argmin_{\bm{c}} \sum_{n\!=\!1}^N\sum_{t\!=\!1}^T \hat{\bm{e}}_t^n(\bm{c})^\intercal\bm{\Gamma}^{n}_t\hat{\bm{e}}_t^n(\bm{c}) + \frac{1}{2}\bm{c}^\intercal\bm{\Sigma}_c^{-1}\bm{c}.
    \end{equation}
    %Since \eqref{eq:obj-reform-f1} is quadratic in $\bm{c}$, it is not directly usable as an objective function because SDP problems require a linear dependence on decision variables as shown in \eqref{eq:generic-sdp-obj}. However, 
    Using the epigraph reformulation \cite{Boyd04}, we equivalently transform \eqref{eq:obj-reform-f1} into a linear optimization problem with quadratic inequality constraints
    \begin{subequations}
        \label{eq:obj-reform-f2}
        \begin{align}
        \label{eq:obj-reform-f2-obj}
        \bm{c^*} = & \argmin_{\bm{c}, q, p} \quad q+p \\
        \label{eq:obj-reform-f2-constr1}
        & \text{ such that}\enspace %\sum_{n\!=\!1}^N\sum_{t\!=\!1}^T \hat{\bm{e}}_t^n(\bm{c})^\intercal\bm{\Gamma}^{n}_t\hat{\bm{e}}_t^n(\bm{c}) 
        \bm{\varepsilon}(\bm{c})\bm{G}\bm{\varepsilon}(\bm{c})^\intercal \leq q \\
        \label{eq:obj-reform-f2-constr2}
        & \hspace{5.2em} \frac{1}{2}\bm{c}^\intercal\bm{\Sigma}_c^{-1}\bm{c}\, \leq p,
        \end{align}
    \end{subequations}
    where $q$ and $p$ are the newly introduced epigraph variables. %In \eqref{eq:obj-reform-f2}, the objective is now a linear term. 
    Thus, it remains to reformulate \eqref{eq:obj-reform-f2-constr1} and \eqref{eq:obj-reform-f2-constr2} as LMI constraints to obtain an SDP.
    % Exploiting the closed-form expression for $\bm{\hat{e}}(\bm{c})$ \eqref{eq:param_error}, constraint \eqref{eq:obj-reform-f2-constr1} can be rewritten using the Schur complement for nonstrict inequalities \cite{zhang05}. To achieve this, we state %introduce $\bm{S}$ and $\bm{R}$ as defined in \eqref{eq:sdp-obj-def-S} and \eqref{eq:sdp-obj-def-R} respectively:
    % \begin{equation}
    %     \label{eq:obj.reform-f3}
    %     \sum_{n\!=\!1}^N\sum_{t\!=\!1}^T \hat{\bm{e}}_t^n(\bm{c})^\intercal\bm{\Gamma}^{n}_t\hat{\bm{e}}_t^n(\bm{c}) = \bm{\varepsilon}(\bm{c})\bm{G}\bm{\varepsilon}(\bm{c})^\intercal,
    % \end{equation}
    % which can be seen directly by inserting \eqref{eq:sdp-obj-def-S} and \eqref{eq:sdp-obj-def-R}. 
    % Thus, constraint \eqref{eq:obj-reform-f2-constr1} becomes
    % %with $\bm{R} \geq 0$ and thus constraint \eqref{eq:obj-reform-f2-constr} becomes:
    % \begin{equation}
    %     \label{eq:obj-reform-f4}
    %     q - \bm{\varepsilon}(\bm{c})\bm{G}\bm{\varepsilon}(\bm{c})^\intercal \geq 0.
    % \end{equation}
    Due to the definition of $\bm{G}$, the properties \mbox{$\text{det}(\bm{G}) \neq 0$} and $\bm{G}=\bm{G}^\intercal$ hold and it follows from the Schur complement \cite{zhang05} that \eqref{eq:obj-reform-f2-constr1} is equivilant to
    \begin{equation}
        \label{eq:obj.reform-schur-q}
        q - \bm{\varepsilon}(\bm{c})\bm{G}\bm{\varepsilon}(\bm{c})^\intercal \geq 0 
        \iff
        \begin{bmatrix}
                q & \bm{\varepsilon}(\bm{c}) \\
                \bm{\varepsilon}(\bm{c})^\intercal & \bm{G}^{-1}
            \end{bmatrix} \succeq 0,
    \end{equation}
    where the positive semidefinite matrix on the right is exactly $\bm{Q}$ \eqref{eq:sdp-obj-def-Q}. Following the same procedure for \eqref{eq:obj-reform-f2-constr2}, we obtain
    \begin{equation}
        \label{eq:obj.reform-schur-p}
        p - \frac{1}{2}\bm{c}^\intercal\bm{\Sigma}_c^{-1}\bm{c} \geq 0 
        \iff
        \begin{bmatrix}
            p & \bm{c}^\intercal \\
            \bm{c} & \bm{\Sigma}_c
        \end{bmatrix} \succeq 0.
    \end{equation}    
    Finally, substituting the quadratic constraints in \eqref{eq:obj-reform-f2} with the right hand side of \eqref{eq:obj.reform-schur-q} and \eqref{eq:obj.reform-schur-p} yields
    % \begin{subequations}
    %     \label{eq:obj-reform-f5}
    %     \begin{align}
    %     \label{eq:obj-reform-f5-obj}
    %     \bm{\hat{\pi}^*} = & \argmin_{\bm{c}, q, p} \quad q+p \\
    %     \label{eq:obj-reform-f5-constr}
    %     & \text{ such that} \begin{bmatrix}
    %             q & \bm{\varepsilon}(\bm{c}) \\
    %             \bm{\varepsilon}(\bm{c})^\intercal & \bm{G}^{-1}
    %         \end{bmatrix} \succeq 0 \\
    %     & \hspace{7em} \!\begin{bmatrix}
    %         p & \bm{c}^\intercal \\
    %         \bm{c} & \bm{\Sigma}_c
    %     \end{bmatrix} \succeq 0,
    %     \end{align}
    % \end{subequations}
    the optimization problem \eqref{eq:sdp-obj} and thus concludes the proof.
\end{prooof}

Solving SDP \eqref{eq:sdp-obj} thus allows us to find a parameterization $\bm{c}^*$ such that the employed control law $\bm{\hat{\pi}}(\bm{c}^*,\bm{x})$ reproduces the observed trajectories. Moreover, inferring $\bm{c}^*$ directly determines $\hat{V}(\bm{c}^*,\bm{x})$ due to \eqref{eq:V_poly}, thereby estimating the underlying value function as well. However, since the SDP problem \eqref{eq:sdp-obj} in its current form does not entail the Lyapunov constraints \eqref{eq:l1b} - \eqref{eq:l1d}, it is not ascertained that the retrieved $\hat{V}(\bm{c}^*,\bm{x})$ is also a CLF and not guaranteed that the employed control law $\bm{\hat{\pi}}(\bm{c}^*,\bm{x})$ asymptotically stabilizes system \eqref{eq:unc_dyn}.  

\subsection{Lyapunov-Constrained Nonlinear SOS Program } \label{subsec:sos-lya-constr}

Parameterizing $\hat{V}(\bm{c},\cdot)$ as a multivariate polynomial %Formulating the search over $\bm{c}$ as an SDP problem 
allows a straightforward extension to introduce the stability constraints \eqref{eq:l1b} - \eqref{eq:l1d} as SOS conditions. To this end, we impose the following assumption on the system dynamics \eqref{eq:unc_dyn}. \looseness=-1
\begin{assumption}
\label{as:poly_dyn}
The functions $\bm{f(\cdot)}$ and $\bm{g(\cdot)}$ in \eqref{eq:unc_dyn} are polynomial.
\end{assumption}
Due to \Cref{as:poly_dyn} the derivative constraint \eqref{eq:l1d} takes a polynomial form and can be expressed using the SOS technique. This is accomplished by decomposing the polynomial into a quadratic form and testing its decomposition matrix for positive semidefiniteness, which is formalized as follows: 
\begin{definition}[\cite{Papachristodoulou05}]
\label{def:sos}
    We call a multivariate polynomial $p(\bm{x})$ of degree $2d$ (with $d \geq 1$) SOS decomposable if and only if there exists a positive semidefinite matrix $\bm{H}$ and a monomial vector $\bm{m}(\bm{x})$ of at least degree $d$ such that
\begin{equation}
            \label{eq:sos-decomp}
    	p(\bm{x}) = \bm{m}(\bm{x})^\intercal\bm{H}\bm{m}(\bm{x}) \quad \text{ s.t. } \bm{H} \succeq 0
\end{equation}
    % \begin{subequations}
    %     \label{eq:sos}
    %     \begin{align}
    %     \label{eq:sos-decomp}
    % 	p(\bm{x}) =& \bm{m}(\bm{x})^\intercal\bm{H}\bm{m}(\bm{x}) \quad \text{ s.t. } \bm{H} \succeq 0\\
    % 	\label{eq:sos-qdecomp}
    % 	& \text{ s.t. } \bm{H} \succeq 0.
    %     \end{align}
    % \end{subequations}
\end{definition}
Any polynomial $p(\bm{x})$ that is SOS decomposable is trivially non-negative. We will denote the set of all polynomials that fulfill \Cref{def:sos} with $\pazocal{S}$. Hence, $p(\bm{x})\in\pazocal{S}$ is equivalent to saying that $p(\bm{x})$ is non-negative.
Since constraints \eqref{eq:l1b} and \eqref{eq:l1d} require strict positivity except at the equilibrium, we offset the Lyapunov conditions by a positive-definite polynomial %$\mu(\bm{x})$
\begin{subequations}
\label{eq:sos-l}
\begin{align}
    \label{eq:sos-l1}
	\hat{V}(\bm{c, x}) - \mu(\bm{x}) &\in \pazocal{S} \\
    \label{eq:sos-l2}
	-\pazocal{L}\hat{V}(\bm{c, x}) - \mu(\bm{x}) &\in \pazocal{S},
\end{align}
\end{subequations}
with $\mu(\bm{x}) > 0, \; \forall \: \bm{x} \ne \bm{0}$. Since $\mu(\cdot)$ can in principle be chosen arbitrarily small, it poses a negligible restriction.
A necessary condition for \eqref{eq:sos-l1} and \eqref{eq:sos-l2} to be SOS decomposable is that the smallest and largest degrees among all monomials in both constraints have to be even. Thus, the coefficients for the smallest and largest degrees among all monomials appearing in both constraints need to be dependent on $\bm{c}$. The following result shows under which conditions this is given.
%In order for \eqref{eq:sos-l1} and \eqref{eq:sos-l2} to pose valid SOS constraints for an optimization problem, it is important to show that a quadratic form is achievable by altering the decision variables $\bm{c}$ (e.g. the coefficients for the smallest and largest degrees among all monomials appearing in both constraints are dependent on $\bm{c}$). This is shown in the following remark.
\begin{lemma}\label{lem:degree_condition}
Consider a control-affine system \eqref{eq:unc_dyn} satisfying \Cref{as:known_dyn} and \ref{as:poly_dyn}, a parameterized polynomial function $\hat{V}(\bm{c},\bm{x})$ according to \eqref{eq:V_poly} and the resulting parametrized control law $\hat{\bm{\pi}}(\bm{c},\bm{x})$ \eqref{eq:c_sontag}. 
%Then the highest and lowest monomial degrees in constraints \eqref{eq:sos-l1} and \eqref{eq:sos-l2} are even and depend on the decision variables $\bm{c}$, 
If the minimum-degree conditions %individual terms obey
\begin{subequations}
\begin{align}
    \mindeg{\lambda(\bm{x})} \wedge \mindeg{\bm{g}(\bm{x})} &= 0, \label{eq:deg-l1} \\
    \mindeg{\bm{f}(\bm{x})} &\geq 1, \label{eq:deg-l2}\\ %\mindeg{\bm{\sigma}}, \mindeg{\bm{f}} &\geq 1, \label{eq:deg-l2}\\
    \mindeg{\mu(\bm{x})} &\geq \mindeg{\hat{V}(\cdot, \bm{x})} = 2, \label{eq:deg-l3}%\\
    %\maxdeg{V(\cdot, \bm{x})} = 2i &\geq \maxdeg{\mu(\bm{x})}, \maxdeg{\bm{f}(\bm{x})}, \label{eq:deg-l4}\\
    %\maxdeg{\lambda(\bm{x})} &= 2j \label{eq:deg-l5}
\end{align}
\end{subequations}
and maximum-degree conditions
\begin{subequations}
\begin{align}
    %\mindeg{\lambda(\bm{x})} \wedge \mindeg{\bm{g}(\bm{x})} &= 0, \label{eq:deg-l1} \\
    %\mindeg{\bm{f}(\bm{x})} &\geq 1, \label{eq:deg-l2}\\ %\mindeg{\bm{\sigma}}, \mindeg{\bm{f}} &\geq 1, \label{eq:deg-l2}\\
    %\mindeg{\mu(\bm{x})} \geq \mindeg{V(\cdot, \bm{x})} &= 2, \label{eq:deg-l3}\\
    \qquad\maxdeg{\hat{V}(\cdot, \bm{x})} &= 2i \geq \maxdeg{\mu(\bm{x})} \wedge \maxdeg{\bm{f}(\bm{x})}, \label{eq:deg-l4}\\
    \maxdeg{\lambda(\bm{x})} &= 2j \label{eq:deg-l5}
\end{align}
\end{subequations}
with ${i, j \in \mathbb{N_+}}$ hold true, then the necessary degree conditions for \eqref{eq:sos-l} are satisfied. %and $\mindeg{\cdot}$ and $\maxdeg{\cdot}$ denoting the lowest and highest degrees of a polynomial's monomials respectively. 
% \begin{gather*}
%     \mindeg{V} = 2, \; \mindeg{\bm{f}} \geq 1, \; \maxdeg{V} = 2i > \maxdeg{\bm{f}}, \; \maxdeg{\lambda} = 2j \; \text{with} \;\; i, j \in \mathbb{N} \\
%     \mindeg{\bm{\sigma}} \geq 1, \; \mindeg{\bm{g}} = 0, \; \mindeg{\lambda} = 0, \; \mindeg{b} \geq \mindeg{V}, \; \maxdeg{b} < \maxdeg{V}
% \end{gather*}
% where $\mindeg{p}$ and $\maxdeg{p}$ denote the degree of the monomial with the lowest and highest degree of all monomials in $p$ respectively. 
\end{lemma}

\begin{prooof}
    Due to \eqref{eq:deg-l3} and \eqref{eq:deg-l4} it follows directly that the lowest and highest monomial degree in \eqref{eq:sos-l1} are even and depend on the decision variables $\bm{c}$.
    %Obeying the degree conditions on $V$ and $b$ from the previous remark, $\mindeg{V}$ and $\maxdeg{V}$ is even and hence the same holds for constraint \eqref{eq:sos-l1}. 
    Thus, it remains to verify the degree conditions for \eqref{eq:sos-l2}. Inserting policy \eqref{eq:c_sontag} yields
    \begin{equation}
        \label{eq:sos-l2-full}
        -\nabla_{x}^\intercal \hat{V}(\bm{c},\bm{x})\bm{f}(\bm{x}) \!\!+\!\! \lambda(\bm{x})\!\norm{\nabla_x^{\intercal} \hat{V}(\bm{c},\bm{x}) \bm{g}(\bm{x})}_{\bm{R}^{-1}} \!\!\!-\mu(\bm{x})\! \in \pazocal{S}.
    \end{equation}
    % \begin{multline}
    %     \label{eq:sos-l2-full}
    %     -\nabla_{x}V^\intercal\bm{f} \; + \; \lambda\nabla_{x}V^\intercal\bm{g}\bm{R}^{-1}\bm{g}^\intercal\nabla_{x}V \\
    %     - \frac{1}{2}\tr\big(\bm{\sigma}^\intercal \nabla_{xx}V\,\bm{\sigma}\big) -\mu \in \pazocal{S},
    % \end{multline}
    %where we omit the dependency on $\bm{x}$ and $\bm{c}$ for improved readability. 
    To verify the degree conditions, we show that the lowest and highest monomial degrees in \eqref{eq:sos-l2-full} are bounded by terms depending on $\hat{V}(\bm{c},\bm{x})$ and are even. The lowest monomial degree of each term in \eqref{eq:sos-l2-full} fulfills 
    \begin{subequations}
        \begin{align}
            \mindeg{ \lambda(\bm{x})\norm{\nabla_x^{\intercal} \hat{V}(\bm{c},\bm{x}) \bm{g}(\bm{x})}_{\bm{R}^{-1}}} & = 2 \label{eq:deg-p1} \\ %{\lambda\nabla_{x}V^\intercal\bm{g}\bm{R}^{-1}\bm{g}^\intercal\nabla_{x}V} & = 2 \label{eq:deg-p1} \\
            \mindeg{-\nabla_{x}^\intercal \hat{V}(\bm{c},\bm{x})\bm{f}(\bm{x})} & \geq 2 \label{eq:deg-p2} \\ %{-\nabla_{x}V^\intercal\bm{f}}  & \geq 2 \label{eq:deg-p2} \\
            % \mindeg{- \tr\big(\bm{\sigma}^\intercal \nabla_{xx}V\,\bm{\sigma}\big)} & \geq 2 \label{eq:deg-p3} \\
            \mindeg{- \mu(\bm{x})} & \geq 2 \label{eq:deg-p3}
        \end{align}
    \end{subequations}
    Thus, due to \eqref{eq:deg-p1}, the lowest possible monomial degree in \eqref{eq:sos-l2-full} is even and depends on $\hat{V}(\bm{c},\bm{x})$. For the highest monomial degree in \eqref{eq:sos-l2-full}, it can be seen that
    \begin{subequations}
        \begin{align}
            \maxdeg{\lambda(\bm{x})\!\norm{\nabla_x^{\intercal} \hat{V}(\bm{c},\bm{x}) \bm{g}(\bm{x})}_{\bm{R}^{-1}}\!} & \!\geq\! \maxdeg{\!-\nabla_{x}^\intercal \hat{V}(\bm{c},\bm{x})\bm{f}(\bm{x})} \label{eq:deg-p11} \\
            \maxdeg{\lambda(\bm{x})\!\norm{\nabla_x^{\intercal} \hat{V}(\bm{c},\bm{x}) \bm{g}(\bm{x})}_{\bm{R}^{-1}}\!} & \!\geq\! \maxdeg{\mu(\bm{x})} \label{eq:deg-p11}
        \end{align}
        % \begin{align}
        %     \maxdeg{\lambda\nabla_{x}V^\intercal\bm{g}\bm{R}^{-1}\bm{g}^\intercal\nabla_{x}V} & \geq \maxdeg{-\nabla_{x}V^\intercal\bm{f}} \label{eq:deg-p11} \\
        %     \maxdeg{\lambda\nabla_{x}V^\intercal\bm{g}\bm{R}^{-1}\bm{g}^\intercal\nabla_{x}V} & \geq \maxdeg{\mu} \label{eq:deg-p11}
        % \end{align}
    \end{subequations}
    due to \eqref{eq:deg-l4}. %    Therefore, it remains to show that the outstanding terms are even. 
    Finally, inserting \eqref{eq:deg-l5} yields
    \begin{align}
        \begin{split}
        \maxdeg{\lambda(\bm{x})\!\norm{\nabla_x^{\intercal} \hat{V}(\bm{c},\bm{x}) \bm{g}(\bm{x})}_{\bm{R}^{-1}}\! }  \!=\! 2(\maxdeg{\hat{V}(\bm{c},\bm{x})}\!\!-\!\!1\!\!+\!\!\maxdeg{\bm{g}(\bm{x})}\!\!+\!\!j), \nonumber %\label{eq:deg-p31} 
        \end{split}
        %{\lambda\nabla_{x}V^\intercal\bm{g}\bm{R}^{-1}\bm{g}^\intercal\nabla_{x}V} & = 2(\maxdeg{V}-1+\maxdeg{\bm{g}} + j). \label{eq:deg-p31}
    \end{align}
    % Furthermore, due to \eqref{eq:deg-l4}, it can be seen that  
    % \begin{equation}
    %     \maxdeg{- \tr\big(\bm{\sigma}^\intercal \nabla_{xx}V\,\bm{\sigma}\big)} = 2\maxdeg{\sigma} + \maxdeg{V} - 2 = 2(\maxdeg{\bm{\sigma}}+i-1). \label{eq:deg-p41}
    % \end{equation}
    % \begin{subequations}
    %     \begin{align}
    %         \maxdeg{\lambda\nabla_{x}V^\intercal\bm{g}\bm{g}^\intercal\nabla_{x}V} & = 2(\maxdeg{V}-1+\maxdeg{\bm{g}}) + \maxdeg{\lambda} \geq \maxdeg{-\nabla_{x}V^\intercal\bm{f}} \notag \\
    %         \maxdeg{- \frac{1}{2}\tr\big(\bm{\sigma}^\intercal \nabla_{xx}V\,\bm{\sigma}\big)} & = 2\maxdeg{\sigma} + \maxdeg{V} - 2 = 2(\maxdeg{\bm{\sigma}}+i-1) \notag
    %     \end{align}
    % \end{subequations}
    % Since \eqref{eq:deg-p31} is even, %both \eqref{eq:deg-p31} and \eqref{eq:deg-p41} are even, 
    which can trivially be seen to be even.
    Hence, the highest monomial degree in \eqref{eq:sos-l2-full} is also guaranteed to be an even and dependent on $\hat{V}(\bm{c},\bm{x})$, which concludes the proof.
    % Which shows that the largest monomial degree is always guaranteed to be an even number dependent on $V$.
    % Finally, we have that the monomial degree of $b$ is bounded by:
    % \begin{equation}
    %     2 \leq \mindeg{b} \leq \maxdeg{b} \leq \maxdeg{\lambda\nabla_{x}V^\intercal\bm{g}\bm{g}^\intercal\nabla_{x}V} \notag
    % \end{equation}
    % which concludes the proof.
\end{prooof}

Intuitively, \eqref{eq:deg-l2}  %and $\bm{\sigma}$ 
ensures vanishing open-loop dynamics $\bm{f}(\cdot)$ %and noise 
at the equilibrium, while \eqref{eq:deg-l1} restricts $\bm{g}(\cdot)$ and $\lambda(\cdot)$ to have non-vanishing control influence of $\nabla_x\hat{V}(\cdot)$ around the equilibrium.
By combining the results of \cref{lem:err_schur} and \ref{lem:degree_condition} we get the SOS-constrained optimization problem \looseness=-1 %By combining \eqref{eq:sdp-obj}, \eqref{eq:sos-l1} and \eqref{eq:sos-l2} we finally get:
\begin{subequations}
    \label{eq:sdp-qmi}
    \begin{align}
        \bm{c^*} = &\argmin_{\bm{c}, q, p} && q + p \\
                    &\text{ such that} && \bm{Q}(q, \bm{c}) \succeq 0,  \\
                    & \phantom{\text{ such that}} && \bm{P}(p, \bm{c}) \succeq 0, \\
        % &\text{such that} && \begin{bmatrix}
        %     q & \bm{S}(\bm{c}) \\
        %     \bm{S}(\bm{c})^\intercal & \bm{R}^{-1}
        % \end{bmatrix} \succeq 0 \\
        \label{eq:sdp-qmi-c1}
        & && \hat{V}(\bm{c, x}) - \mu(\bm{x}) \in \pazocal{S}, \\
        \label{eq:sdp-qmi-c2}
        & && -\pazocal{L}\hat{V}(\bm{c, x}) - \mu(\bm{x}) \in \pazocal{S}.
    \end{align}
\end{subequations}
Optimizing \eqref{eq:sdp-qmi} over $\bm{c}$ yields an estimate $\hat{V}(\bm{c}^*,\bm{x})$ that fulfills the Lyapunov constraints \eqref{eq:l1b} - \eqref{eq:l1d} while also encoding agent preferences through the parameterized log-likelihood \eqref{eq:param_loglik}. 
Here, $\hat{V}(\bm{c}^*,\bm{x})$ concurrently determines the policy $\bm{\hat{\pi}}(\bm{c}^*,\bm{x})$ through its gradient $\nabla_x \hat{V}(\cdot)$ in \eqref{eq:c_sontag}, which is guaranteed to asymptotically stabilize system \eqref{eq:unc_dyn} due to the SOS constraints \eqref{eq:sdp-qmi-c1} and \eqref{eq:sdp-qmi-c2}.
% Moreover, $\bm{c}$ directly determines the policy $\bm{\hat{\pi}}$ \eqref{eq:c_sontag}, which is guaranteed to asymptotically stabilize system \eqref{eq:unc_dyn}, which is certified by $V$. 
Thus, solving \eqref{eq:sdp-qmi} constitutes an integrated certification and stabilization framework. \looseness=-1

However, \eqref{eq:sdp-qmi-c2} remains a restriction preventing the direct implementation of \eqref{eq:sdp-qmi} as a convex SDP problem, since the constraint is nonlinear in the decision variables $\bm{c}$. This can be seen when writing out \eqref{eq:sdp-qmi-c2} in the parameterized form
\begin{equation}
    \label{eq:sos-l2-problem-structure}
    -\bm{c}\nabla_x^{\intercal} \bm{z}(\bm{x})\bm{f}(\bm{x}) +\! \lambda(\bm{x})\!\norm{\bm{c}\nabla_x^{\intercal} \bm{z}(\bm{x}) \bm{g}(\bm{x})}_{\bm{R}^{-1}} \!-\!\mu(\bm{x})\! \in \!\pazocal{S},
% \begin{split}
%     \label{eq:sos-l2-problem-structure}
%     -\bm{c}\nabla_{x}\bm{Z}^\intercal\bm{f} +  \lambda\big(\bm{c}\,&\nabla_{x}\bm{Z}^\intercal\bm{g}\bm{R}^{-1}\bm{g}^\intercal\nabla_{x}\bm{Z}\,\bm{c}^\intercal\big) \\ &\;\;- \frac{1}{2}\tr\big(\bm{\sigma}^\intercal \nabla_{xx}V(\bm{c})\bm{\sigma}\big) -\mu \in \pazocal{S},
% \end{split}
\end{equation}
where the second term is quadratic in $\bm{c}$, thereby yielding a quadratic matrix inequality (QMI). 
% \subsection{Convexification of Constrained Optimization} \label{subsec:sos-convexify}
% \textcolor{cyan}{As result \eqref{eq:sdp-qmi} of the previous section, we derived an SOS-based formulation of the Lyapunov constraints \eqref{eq:l1b} - \eqref{eq:l1d} along with the approximate log-likelihood \eqref{eq:approx_loglik}, which infers a Lyapunov function $V$ from human demonstrations. In extension, this Lyapunov function also represents a value function for some meaningful cost, as established earlier by the property of inverse optimality.} Consider again constraint \eqref{eq:sdp-qmi-c2}:
% \begin{equation}
%     \label{eq:sos-l2-problem-structure}
%     -\bm{c}^\intercal\nabla_{x}\bm{Z}\big(\bm{f} - \lambda\bm{g}\bm{g}^\intercal\nabla_{x}\bm{Z}^\intercal\bm{c}\big) - \frac{1}{2}\tr\big(\bm{\sigma}^\intercal \nabla_{xx}V(\bm{c})\,\bm{\sigma}\big) -b \in \pazocal{S}
% \end{equation}
% The second term in \eqref{eq:sos-l2-problem-structure} is nonlinear in the decision variables $\bm{c}$ and therefore does not adhere to the structure of an SDP problem. Here, the resulting matrix inequality is quadratic in $\bm{c}$, yielding a quadratic matrix inequality (QMI). 
While some QMI define a convex solution set and allow an equivalent reformulation as an LMI \cite{Wang16}, this not the case for \eqref{eq:sos-l2-problem-structure}, which is demonstrated in the following proposition.
%A QMI can define a convex set \myref{[Wang16]} but the problem at hand is in general nonconvex and does not allow an equivalent reformulation as an LMI. This is shown in the following proposition:
\begin{proposition}
    The solution set of \eqref{eq:sos-l2-problem-structure} is nonconvex and cannot be reformulated to an equivalent LMI constraint. \looseness=-1
\end{proposition}
\begin{prooof}
    \label{prop:qmi-nonconvex}
    Consider the linear one dimensional system $f(x) = x$ with $g(x) = 1$ and $x \in \mathbb{R}^1$. Further %assume a deterministic setting with $\sigma = 0$ and 
    choose a quadratic Lyapunov candidate $V(c,x) = cx^2$. Then \eqref{eq:sos-l2-problem-structure} becomes
    \begin{equation}
        \label{eq:ex1:step1}
        -2cx\big(x - \lambda2cx\big) - \mu(x) \in \pazocal{S},
    \end{equation}
    with a bounding polynomial $\mu(x) := wx^2$ for some small \mbox{$1 \gg w > 0$}. Thus, \eqref{eq:ex1:step1} can be rewritten as
    \begin{equation}
        \label{eq:ex1:step2}
        x\big(4\lambda c^2 - 2c - w\big)x \in \pazocal{S}.
    \end{equation}
    It can be seen directly that the SOS condition holds true, if
    %By choosing the monomial vector for the SOS decomposition as $\bm{M}(x) = x$, one can directly infer the LMI of the SOS constraint:
    \begin{equation}
        \label{eq:ex1:step3}
        4\lambda c^2 - 2c - w \geq 0.
    \end{equation}
    Without loss of generality, choose $\lambda=\frac{1}{2}$ and exploit the fact that $w$ can be chosen arbitrarily close to $0$ to express \eqref{eq:ex1:step3} as a strict inequality
    \begin{equation}
        \label{eq:ex1:step4}
        2c^2 > 2c.
    \end{equation}
    It is straight forward to infer the solution set from \eqref{eq:ex1:step4} to
    \begin{equation}
        \{c \in \mathbb{R} \mid c > 1 \; \cup \; c < 0 \},
    \end{equation}
    which is nonconvex, thereby concluding the proof by counter example. %This shows that the quadratic terms in the second Lyapunov constraint \eqref{eq:sos-l2-problem-structure} cannot be rewritten equivalently into a convex constraint by using the Schur complement.
\end{prooof}

Thus, the constrained optimization problem \eqref{eq:sdp-qmi} requires solving a QMI due to \eqref{eq:sdp-qmi-c2}. While there exist approaches to solve optimization problems with nonlinear matrix inequalities \cite{Goh95, Lasserre01, Dinh12}, they are typically prohibitively expensive for larger problems or alter the accompanying objective function. Therefore, in the following section we propose a procedure to solve \eqref{eq:sdp-qmi} in a convex form.

\subsection{Convexified Lyapunov-Constrained Optimization} \label{subsec:sos-convexify}
If a QMI is bilinear \cite{Wang16}, it can be solved efficiently using a convexification approach.
%Given a particular form of a QMI, namely a bilinear form \cite{Wang16}, the problem can be solved efficiently using a convexification approach. 
% Such a bilinear matrix inequality (BMI) is a special case of a QMI \myref{[Wang16]} and takes the general structure
% \begin{equation}
%     \label{eq:def-bmi}
%     \bm{B}(\bm{c}, \bm{d}) = \bm{F_0} + \sum_{i\!=\!1}^p c_i\bm{F_i} + \sum_{i\!=\!1}^q d_i\bm{G_i} + \sum_{i\!=\!1}^p\sum_{j\!=\!1}^q c_id_j\bm{F_{i, j}} \succeq 0,
% \end{equation}
% with symmetric real matrices $\bm{F_i}$, $\bm{G_i}$ and $\bm{F_{i, j}}$. 
% Here, the bilinear form can be exploited to reduce \eqref{eq:def-bmi} to a convex LMI problem by optimizing over only one of the decision variables. Thus, optimizing alternately between the two parameter sets yields a sequence of convex SDPs, which is exploited by so-called alternating approaches to solve BMIs \myref{[Goh94]}. 
In particular, the bilinear form can be exploited to reduce a QMI to a convex LMI by fixing one set of decision variables and optimizing over the remaining ones. Thus, optimizing alternately between the two parameter sets yields a sequence of convex SDPs \cite{Goh94}.
However, employing this scheme directly to \eqref{eq:sdp-qmi} is not possible, since the constraint \eqref{eq:sdp-qmi-c2} is not bilinear. % and therefore the approach is not guaranteed to yield a feasible solution. 
To overcome this, %we propose to apply the alternating approach to a proxy constraint to generate a convex inner approximation of the constrained optimization problem \eqref{eq:sdp-qmi}. %Thereby we retain the computational efficiency, whilst guaranteeing that the stability constraints are fulfilled. 
we make use of the structure of the derivative constraint \eqref{eq:sos-l2-problem-structure}. In particular, under mild conditions it is always possible to find a feasible solution to the Lyapunov-constrained optimization \eqref{eq:sdp-qmi} using a sufficiently large ${\lambda(\cdot)}$, which is demonstrated in the following. \looseness=-1

\begin{lemma}
    \label{lem:lya-synthesis}
    Consider a control-affine, continuous system \eqref{eq:unc_dyn} satisfying the degree conditions of \Cref{lem:degree_condition}. If the set
    \begin{equation}
        \label{eq:assumption-control}
        \{ \bm{x} \mid \bm{g}(\bm{x}) = \bm{0}, \nabla_{x}^\intercal V(\bm{x})\bm{f}(\bm{x}) \geq 0, 
         \bm{x} \ne \bm{0} \} %\bm{\sigma}(\bm{x}) > \bm{0},
    \end{equation}
    is empty and a Lyapunov function candidate $V(\bm{x})$ satisfying
    \begin{align}
        \label{eq:prop:lya-root0}
        V(\bm{x}) > 0, \quad \forall \bm{x} \ne \bm{0} \\
        \label{eq:prop:lya-root}
        \nabla_{x}V(\bm{x}) \ne \bm{0}, \quad \forall \bm{x} \ne \bm{0}
    \end{align}
    is given, there always exists a continuous, positive definite $\lambda(\bm{x})$ that asymptotically stabilizes \eqref{eq:unc_dyn} under the stabilizing control policy $\bm{\hat{\pi}}(\bm{x})$ \eqref{eq:sontag}.
    % \begin{equation}
    %     {\bm{u}}(\bm{x}) = -\lambda(\bm{x})\bm{g}(\bm{x})^\intercal\nabla_{x}V(\bm{x}). \label{eq:lem:pol}
    % \end{equation} %where $V$ is a CLF.
\end{lemma}

\begin{prooof}
    %Due to \eqref{eq:prop:lya-root0}, the first Lyapunov constraint is fulfilled by definition. The second Lyapunov constraint under the control law \eqref{eq:lem:pol} after rearranging yields
    Since $V(\cdot)$ is positive definite due to \eqref{eq:prop:lya-root0}, it remains to show that $\dot{V}(\cdot)$ is negative definite. After inserting control law \eqref{eq:sontag} and rearranging we obtain
    \begin{equation}
        \label{eq:proof:l2-arranged}
        \lambda(\bm{x})\norm{\nabla_x^{\intercal} V(\bm{x}) \bm{g}(\bm{x})}_{\bm{R}^{-1}} > \nabla_{x}^\intercal V(\bm{x})\bm{f}(\bm{x}) \quad \forall \bm{x} \ne \bm{0}
    \end{equation}
    % \begin{multline}
    %     \label{eq:proof:l2-arranged}
    %     \lambda(\bm{x})\big(\bm{g}(\bm{x})^\intercal\nabla_{x}V(\bm{x}))^2 > \\
    %     \nabla_{x}V(\bm{x})^\intercal\bm{f}(\bm{x}) + \frac{1}{2}\tr\big(\bm{\sigma}^\intercal \nabla_{xx}V(\bm{c})\,\bm{\sigma}\big) \quad \forall \bm{x} \ne \bm{0}
    % \end{multline}
    For any state with $\bm{g}(\bm{x}) = \bm{0}$, the right hand side of \eqref{eq:proof:l2-arranged} needs to be negative, which is guaranteed since the set \eqref{eq:assumption-control} is empty. Otherwise, we can always choose a sufficiently large $\lambda(\bm{x}) > 0$ such that \eqref{eq:proof:l2-arranged} is fulfilled, since the norm term on the left is guaranteed to be positive definite due to \eqref{eq:prop:lya-root} and $\bm{R}$ being positive definite, which concludes the proof.
\end{prooof}

Intuitively, \eqref{eq:assumption-control} requires that the system needs to be controllable in states where the stability constraints are violated by the open-loop dynamics, which is not particularly restrictive. Moreover, \cref{lem:lya-synthesis} states that a control policy that follows the direction of $\nabla_x V(\cdot)$ %, such as \eqref{eq:lem:pol} or Sontag's formula \eqref{eq:c_sontag}, 
is guaranteed to exist and stabilize the system given a sufficiently large ${\lambda}(\cdot)$, if $V(\cdot)$ has a non-zero gradient everywhere except at the equilibrium. Therefore, we propose to substitute the derivative constraint \eqref{eq:sdp-qmi-c2} by imposing a non-zero gradient constraint \eqref{eq:prop:lya-root} on $\hat{V}(\cdot,\cdot)$ and additionally optimize over ${\lambda}(\cdot)$. 
% In other words, \cref{lem:lya-synthesis} states that we can always stabilize the system if $\bm{g}(\bm{x}) \ne \bm{0}$. Since our control is determined by $\nabla_{x}V$, we need to enforce $\nabla_{x}V \ne 0$ in order to drive our system in states where the open loop dynamics are unstable.
%However, a sufficient condition for the existence of such a ${\lambda}$ is a non-zero gradient of $V$ for all states except the equilibrium point \eqref{eq:prop:lya-root}, which is currently not enforced. 
In order to avoid restricting the solution space unnecessarily, e.g., by enforcing convexity, we propose an iterative procedure, where the solution at each iteration $i$ is guaranteed to yield a $\hat{V}(\bm{c}_i,\cdot)$ that satisfies the non-zero gradient condition \eqref{eq:prop:lya-root}. % by leveraging the SOS framework. %and solving an SDP problem, as shown in the following.
%To our best knowledge, there is no straight-forward way to solve this constraint without further restrictions, however, it is possible to find an approximate solution that is guaranteed to satisfy this constraint by solving an SDP problem, as shown in the following lemma:

\begin{lemma}
    \label{lem:gs}
    Given a parameter vector $\bm{c}_i$ at iteration $i$ for which a polynomial $\hat{V}(\bm{c}_i, \bm{x})$ of the form \eqref{eq:V_poly} satisfies
    \begin{equation}
        \label{eq:prop:gradsquare}
        \nabla_{x}^\intercal \hat{V}(\bm{c}_i, \bm{x})\nabla_{x}\hat{V}(\bm{c}_i, \bm{x}) > 0, \quad \forall \bm{x} \ne \bm{0},
    \end{equation}
    then solving the SOS-problem %for $\bm{c}_{i+1}$
    \begin{align}
        \label{eq:prop:gradsquare-sos}
        \bm{c}_{i+1} = \text{find }& \enspace \bm{c} \\
        \text{s.t. }& \; \nabla_{x}^\intercal \hat{V}(\bm{c}_i, \bm{x})\nabla_{x}\hat{V}(\bm{c}, \bm{x}) - \mu(\bm{x}) \in \pazocal{S},
    \end{align}
    at iteration $i\!+\!1$ yields a parameter vector $\bm{c}_{i+1}$ for which $\hat{V}(\bm{c}_{i+1}, \bm{x})$  fulfills the non-zero gradient constraint \eqref{eq:prop:lya-root}.
\end{lemma}

\begin{prooof}
    %Given a feasible initial guess $\bm{c}_i$ that satisfies \eqref{eq:prop:gradsquare}, we solve the SOS-problem \eqref{eq:prop:gradsquare-sos} which yields the coefficients $\bm{c}_{i+1}$. 
    Since $\mu(\bm{x})$ is positive definite, we get for \eqref{eq:prop:gradsquare-sos}
    %Using the fact that $\mu(\bm{x}) > 0 \; \forall \bm{x} \ne \bm{0}$ and SOS implying non-negativity, we get
    \begin{equation}
        \label{eq:proof:gradsquare-sos}
        \nabla_{x}^\intercal \hat{V}(\bm{c}_i, \bm{x})\nabla_{x}\hat{V}(\bm{c}_{i+1}, \bm{x}) > 0 \quad \forall \bm{x} \ne \bm{0}.
    \end{equation}
    Squaring both sides of \eqref{eq:proof:gradsquare-sos} yields
    \begin{equation}
        \nabla_{x}^\intercal \hat{V}_{i}(\bm{x})\nabla_{x}\hat{V}_{i+1}(\bm{x})\nabla_{x}^\intercal \hat{V}_{i}(\bm{x})\nabla_{x}\hat{V}_{i+1}(\bm{x}) > 0 \\ \quad \forall \bm{x} \ne \bm{0}, \nonumber
    \end{equation}
    where $\hat{V}_{i}(\bm{x}) := \hat{V}(\bm{c}_i, \bm{x})$ and $\hat{V}_{i+1}(\bm{x}) := \hat{V}(\bm{c}_{i+1}, \bm{x})$. This can only hold if $\nabla_{x}\hat{V}_{i+1}(\bm{x})$ is not zero outside of the origin, i.e., fulfills \eqref{eq:prop:lya-root}, since it is already given that $\nabla_{x}\hat{V}_{i}(\bm{x})$ satisfies \eqref{eq:prop:lya-root} due to \eqref{eq:prop:gradsquare}, which concludes the proof.
\end{prooof}

\renewcommand{\algorithmicrequire}{\textbf{Input:}}
\begin{algorithm}[t]
\small
\caption{Control Lyapunov Landscapes (CLL) - IRL}\label{alg:cap}
\begin{algorithmic}[1]
\Require $\bm{c}_0$, $\bm{\theta}_0$, $D = \{(\bm{x}^n_t)\}^T_{t=0}$
\Ensure $\bm{c}_0 = \{ \hat{V}(\bm{c}_0, x) \mid \hat{V}(\bm{c}_0, x) \in \pazocal{S}, \hat{V}(\bm{c}_0, x) \text{ s.t. \eqref{eq:sdp-gs}} \}$
\Ensure $\bm{\theta}_0 = \{ \lambda(\bm{\theta}_0, x) \mid \lambda(\bm{\theta}_0, x) > 0 \quad \forall \bm{x} \neq \bm{0} \}$
\For{$i$= $0 \dots \textit{\#Iterations}$}
    \State $\bm{c}_{i+1} \gets $ solve convex SDP \eqref{eq:sdp-gs} initialized with $\bm{c} = \bm{c_i}$
\EndFor
\State $\bm{c}^* \gets \bm{c_i}$ with $i = \argmin \{ q_i + p_i\}$
\State $\bm{\theta}^* \gets$ solve convex SDP \eqref{eq:sdp-lambda}
\State $\hat{V} \gets \hat{V}(\bm{c}^*, \bm{x})$
\State $\hat{\pi} \gets$ \eqref{eq:final_sontag} with $\bm{c} = \bm{c}^*$, $\bm{\theta} = \bm{\theta}^*$
\State \Return $\hat{V}$, $\hat{\pi}$
\end{algorithmic}
\label{alg:sequence}
\end{algorithm}

Note that the application of \cref{lem:gs} in an iterative procedures requires an initial feasible guess {$\bm{c}_0$} that fulfills \eqref{eq:prop:gradsquare}, which can be trivially chosen as ${\hat{V}(\bm{c}_0, x) = \sum_{i\!=\!1}^n x_i^2}$. Combining the results of Lemma \ref{lem:lya-synthesis} and Lemma \ref{lem:gs}, it is possible to construct a sequence of convex SDP problems which yields a solution to the original Lyapunov-constrained optimization problem \eqref{eq:sdp-qmi}. In particular, we first deploy the constraints in \cref{lem:gs} to iteratively find Lyapunov function candidates with non-vanishing gradients, then we optimize over $\lambda(\cdot)$ to ensure that $\pazocal{L}\hat{V}(\cdot,\cdot)$ is negative definite. \Cref{alg:sequence} explains the programmatic structure for solving the convexified Lyapunov-constrained optimization problem, while \Cref{th:sequential} proves the adherence to the stability constraints.

\begin{theorem} \label{th:sequential}
    Consider a control-affine system \eqref{eq:unc_dyn} satisfying \Cref{as:known_dyn} and \ref{as:poly_dyn}, which adheres to the degree conditions in \Cref{lem:degree_condition}, %\textcolor{red}{with $\bm{f}(\bm{0}) = \bm{0}$} 
    and an empty set \eqref{eq:assumption-control} along with a control policy %$\hat{\bm{\pi}}(\bm{\theta, c, x})$
    \begin{equation}
        \begin{split}
        \hat{\bm{\pi}}&(\bm{\theta, c, x}) \!= \!\! \\
        &\enspace\,\begin{cases}
            -\lambda(\bm{\theta}, \bm{x})\bm{R}^{-1}\bm{g}(\bm{x})^\intercal  \nabla_x \hat{V}(\bm{c},\bm{x}), & \! \bm{g}^\intercal\nabla_x \hat{V} \neq 0\\
            \bm{0},              & \! \bm{g}^\intercal\nabla_x \hat{V} = 0,
        \end{cases}
        \end{split}
     \label{eq:final_sontag}
    \end{equation}    
    where $\lambda(\bm{\theta}, \bm{x})$ and $\hat{V}(\bm{c}, \bm{x})$ are polynomials parameterized by $\bm{\theta}$ and $\bm{c}$ respectively. If an initial parameterization %$\bm{c}_0$ for which $V(\bm{c}_0, \bm{x})$ fulfills \eqref{eq:prop:gradsquare} 
    $\{\bm{\theta}_0, \bm{c}_0\}$ for which $\lambda(\bm{\theta}_0, \bm{x})$ is positive definite and $\hat{V}(\bm{c}_0, \bm{x})$ fulfills \eqref{eq:prop:gradsquare}  %${\lambda(\bm{\theta}_0, \bm{x}) > 0, \; \forall \bm{x} \ne \bm{0}}$ and ${\nabla_{x}V(\bm{c}_0, \bm{x})^\intercal\nabla_{x}V(\bm{c}_0, \bm{x}) > 0, \; \forall \bm{x} \ne \bm{0}}$ 
    is given, then solving the SDP problems
    \begin{subequations}
        \label{eq:sdp-gs}
        \begin{align}
            {\bm{c}}^* = &\argmin_{\bm{c}, q, p} && q + p\\
            % &\text{such that} && \begin{bmatrix}
            %     q & \bm{S}(\bm{a}_0, \bm{c}) \\
            %     \bm{S}(\bm{a}_0, \bm{c})^\intercal & \bm{R}^{-1}
            % \end{bmatrix} \succeq 0 \\
              &\text{ s.t.} && \bm{Q}(q, \bm{\theta}_0, \bm{c}) \succeq 0,  \\
              & \phantom{\text{ s.t.}} && \bm{P}(p, \bm{c}) \succeq 0, \\
            \label{eq:sdp-gs-c1}
            & && \hat{V}(\bm{c, x}) - \mu(\bm{x}) \in \pazocal{S} \\
            \label{eq:sdp-gs-c2}
            & && \nabla_{x}^\intercal \hat{V}(\bm{c}_0, \bm{x})\nabla_{x}\hat{V}(\bm{c}, \bm{x}) - \mu(\bm{x}) \in \pazocal{S}
        \end{align}
    \end{subequations}
    to obtain a parameterized $\hat{V}(\bm{c}^*, \bm{x})$ and subsequently solving
    \begin{subequations}
        \label{eq:sdp-lambda}
        \begin{align}
            {\bm{\theta}}^* = &\argmin_{\bm{\theta}, q, p} && q + p \\
            % &\text{such that} && \begin{bmatrix}
            %     q & \bm{S}(\bm{a}, {\bm{c}}^*) \\
            %     \bm{S}(\bm{a}, {\bm{c}}^*)^\intercal & \bm{R}^{-1}
            % \end{bmatrix} \succeq 0 \\
              &\text{ s.t.} && \bm{Q}(q, \bm{\theta}, \bm{c}^*) \succeq 0,  \\
              & \phantom{\text{ s.t.}} && \bm{P}(p, \bm{c}^*) \succeq 0, \\
            \label{eq:sdp-lambda-c1}
            & && \lambda(\bm{\theta}, \bm{x}) \in \pazocal{S} \\
            \label{eq:sdp-lambda-c2}
            & && -\pazocal{L}\hat{V}(\bm{\theta}, {\bm{c}}^*, \bm{x}) - \mu(\bm{x}) \in \pazocal{S}
        \end{align}
    \end{subequations}
    yields a policy $\hat{\bm{\pi}}(\bm{\theta}^*, \bm{c}^*,\bm{x})$ that renders system \eqref{eq:unc_dyn} asymptotically stable with $\hat{V}(\bm{c}^*, \bm{x})$ satisfying the Lyapunov constraints \eqref{eq:l1b} - \eqref{eq:l1d}. 
\end{theorem}

\begin{prooof}
    The positive definiteness condition of the first Lyapunov constraint \eqref{eq:l1b} is ensured by \eqref{eq:sdp-gs-c1}. Furthermore, due to Lemma \ref{lem:gs}, constraint \eqref{eq:sdp-gs-c2} guarantees that the gradient $\nabla_{x}\hat{V}(\bm{c}^*, \bm{x})$ satisfies \eqref{eq:prop:lya-root}. %, while \eqref{eq:sdp-gs-c1} is equivalent to \eqref{eq:sdp-qmi-c1}. 
    Consequently, this ensures that the second SDP problem \eqref{eq:sdp-lambda} has a non-empty solution set, i.e., that a positive and sufficiently large $\lambda(\bm{\theta}, \bm{x})$ that satisfies \eqref{eq:sdp-lambda-c2} exists, as stated by \cref{lem:lya-synthesis}. By imposing the SOS condition \eqref{eq:sdp-lambda-c1}, the non-negativity of $\lambda(\bm{\theta}, \bm{x})$ is ensured. Thus, solving the SDP problems \eqref{eq:sdp-gs} and \eqref{eq:sdp-lambda} sequentially is guaranteed to yield a parametrization $\{{\bm{\theta}}^*, {\bm{c}}^*\}$ that fulfills the Lyapunov conditions \eqref{eq:sdp-gs-c1} and \eqref{eq:sdp-lambda-c2}, which concludes the proof.
\end{prooof}

\begin{figure}
    \centering
    \includegraphics[width=.41\textwidth]{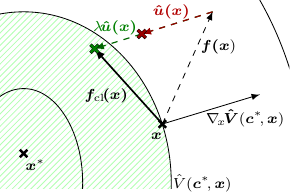}
    \caption{Geometric interpretation of the sequential optimization in \cref{alg:sequence}. In red $\hat{\bm{u}}$ shows the control input after only optimizing parameter vector $\bm{c}$, while $\lambda\hat{\bm{u}}$ in green represents the resulting control input after optimizing $\bm{\theta}$. If the conditions in \cref{lem:lya-synthesis} are met, a sufficiently large $\lambda(\bm{\theta},\cdot)$ s.t. the policy $\bm{\hat{\pi}}(\bm{\theta}, \bm{c}, \cdot)$ induced by $\hat{V}(\bm{c}, \cdot)$ stabilizes the closed-loop system $\bm{f}_{\text{cl}}(\cdot)$ is found.}
    \label{fig:gs:geometric}
\end{figure}

Intuitively, the first optimization \eqref{eq:sdp-gs} finds a candidate $\hat{V}(\bm{c}^*, \bm{x})$ and the second problem \eqref{eq:sdp-lambda} scales the resulting control law $\hat{\bm{\pi}}(\bm{\theta}^*, \bm{c}^*,\bm{x})$ such that $\hat{V}(\bm{c}^*, \bm{x})$ is a valid CLF for the system. This is conceptually illustrated in \Cref{fig:gs:geometric}. While sequentially solving the SDP problems \eqref{eq:sdp-gs} and \eqref{eq:sdp-lambda} is computationally efficient, the proxy constraint \eqref{eq:sdp-gs-c2} leads to a more conservative solutions, as it is a convex inner approximation of the original problem \eqref{eq:sdp-qmi}. However, 
differently to other shape-constrained polynomial regression concepts \cite{Curmei20}, the approach in \cref{th:sequential} is less restrictive, since it does not impose strict monotonicity of the gradient of $\hat{V}(\bm{c}^*, \bm{x})$. %Moreover, since each optimization step of the proposed algorithm is convex, convergence to a unique solution is guaranteed, which facilitates 
% the integrated certification and stabilization framework we propose.

\section{Simulation Evaluation}\label{sec5}
For the evaluation of the proposed approach, we consider two settings. First, we illustrate the inverse optimality of the learned CLF in a scenario where the ground truth value function is known. Here, we show for a nonlinear system that our algorithm is able to recover $V^*(\cdot)$ up to a small error.

\subsection{Setup and Sample Generation}
\begin{figure}
    \centering
    \includegraphics[width=.48\textwidth]{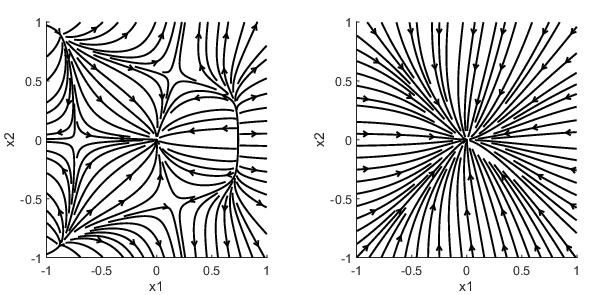}
    % \begin{minipage}[t]{0.22\textwidth}
    %     \centering
    %     \includegraphics[width=\textwidth]{pdfs/test_figures/ground_truth/gt_open_loop.pdf}
    %     \subcaption{\label{fig:eval:GT:openloop}}
    % \end{minipage}
    % \hfill
    % \begin{minipage}[t]{0.22\textwidth}
    %     \centering
    %     \includegraphics[width=\textwidth]{pdfs/test_figures/ground_truth/gt_closed_loop.pdf}
    %     \subcaption{\label{fig:eval:GT:closedloop}}
    % \end{minipage}
    \caption{Visualization of the system dynamics. (Left) open-loop dynamics $\bm{f}$ and (right) closed-loop dynamics $\bm{f^*}$ under $\bm{\pi}^*$.}
    \label{fig:eval:GT:dynamics}
\end{figure}

In order to generate a problem setting where the value function $V^*(\cdot)$ is known, we employ the converse HJB (CoHJB) \cite{Doyle96}. Given a value function $V^*(\cdot)$ and the dynamics component $\bm{g}(\cdot)$, the HJB equation becomes linear in the remaining terms $\bm{f}(\cdot)$, $l(\cdot)$ and $r(\cdot)$. Thus, the CoHJB yields the system dynamics for which a given $V^*(\cdot)$ is the value function under the specified cost function. For the polynomial case, % solving the HJB via the CoHJB method 
the problem reduces to a system of equations in the unknown coefficients of $\bm{f}(\cdot)$. Consider the following nonlinear value function
\begin{equation}
    V^* = 0.1 x_1^2 + 0.5 x_2^2 + x_1 x_2^2 + x_1^4 + x_2^4,
\end{equation}
with states $\bm{x} := [x_1, x_2]^\intercal$ and input matrix $\bm{g}(\bm{x}) = I_2$. Solving the CoHJB problem, we get the open loop system dynamics
\begin{equation}
    \bm{f}(x) = \begin{bmatrix}
                    -0.5477 x_1 + 0.25 x_2^2 + x_1^3 \\
                    -0.3672 x_2 + 0.5 x_1 x_2 + x_2^3,
                \end{bmatrix}
\end{equation}
which are shown in \Cref{fig:eval:GT:dynamics} (left). Since $V^*(\cdot)$ is known, we can directly compute the optimal policy $\bm{\pi}^*(\cdot)$ using \eqref{eq:opt_feedback}, thus, providing the closed-loop dynamics $\bm{f}^*(\cdot)$, which are illustrated in \Cref{fig:eval:GT:dynamics} (right). By introducing state-dependent noise 
\begin{equation}
    \label{eq:eval:GT:variance}
    \bm{\sigma}(\bm{x}) = \begin{bmatrix}
        0.001 \\
        0.001
    \end{bmatrix} (x_1^2 + x_2^2),
\end{equation}
we further obtain the stochastic closed-loop dynamics \eqref{eq:cl_dyn_2} from which demonstrations can be sampled.
In order to learn the CLF, we observe one-step trajectories of the stochastic closed-loop system dynamics %, i.e., the data set consists of observation tuples $\big(\bm{x_i}, \bar{\bm{f}}(\bm{x_i}) + \bm{\sigma}(\bm{x_i}) \big)$, 
over a grid spanning different parts of the state space. 
%We start by investigating the ability to restore the original value function $V^*$ for a problem with a 2D statespace from observations sampled over a 2D grid. 
%While grid-based sampling is not representative for most learning from demonstrations settings, it highlights the performance of the learning methods when data is available and contrastingly shows the generalization properties in data-free areas of the state space. % impact of having data over a broader state space and the structure that polynomials and SOS-constraints impose for data-free areas. 
Hence the demonstrations consist of observation tuples $\big(\bm{x_i}, (\bm{f}^*(\bm{x_i}) + \bm{\sigma}(\bm{x_i}))\Delta t \big)$. Without loss of generality the measurements are normalized, thus, focusing on the state space $\pazocal{X} \in [-1, 1]^2$. % to avoid numerical problems for high-order polynomials.
Since the true value function $V^*(\cdot)$ is typically unknown, an expressive parameterization for the candidate $\hat{V}(\bm{c},\cdot)$ is chosen, which provides more flexibility during the inference. Here, a degree $d\!=\!8$ is selected, while the true value function is of degree $d^*\!=\!4$. Thereby the principle capacity to find $V^*(\cdot)$ exists, however, it is required to search over a substantially larger parameter space. As an initial guess we choose $\hat{V}_0 = 0.1 \big(x_1^2 + x_2^2\big)$ and $\lambda_0(x) = 0.5$. To solve the SDP problems we employ the YALMIP Matlab toolbox \cite{Lofberg04} together with the MOSEK SDP solver. \looseness=-1 %\textcolor{red}{and utilize YALMIPs built-in SOS processing routines \myref{[Lofberg09]}}. 

\begin{figure*}
    \centering
    \includegraphics[width=.925\textwidth]{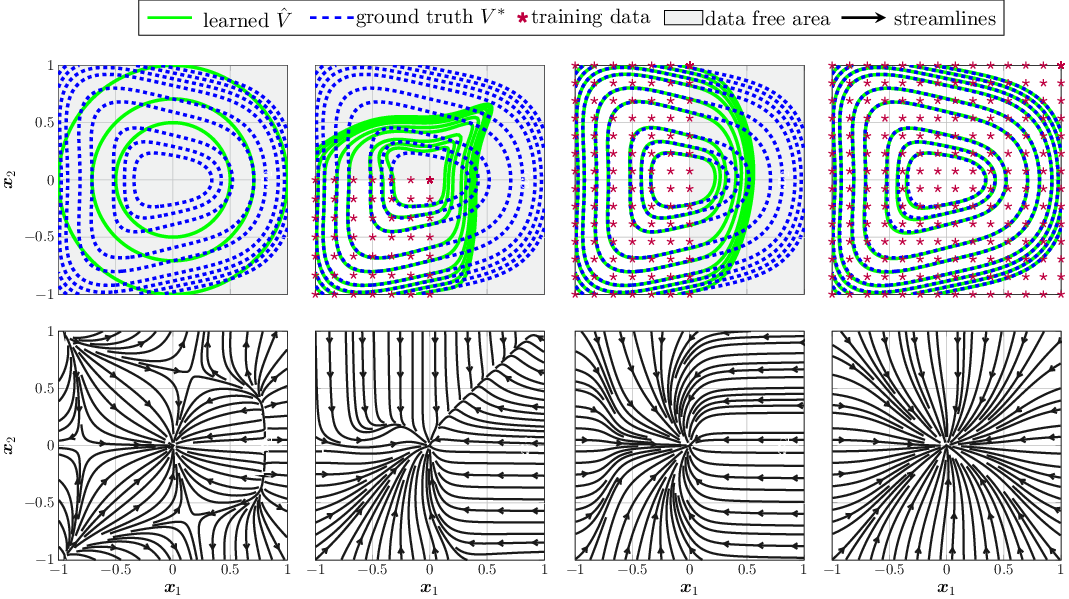}
    \caption{Ground truth evaluation of the proposed algorithm. The top row shows the contours of the ground truth value function $V^*(\cdot)$ (blue, dashed) and the learned CLF $\hat{V}(\cdot, \cdot)$ (green). Training data is depicted with purple stars, while the greyed out parts show data free areas. The bottom row illustrates the learned closed-loop dynamics. The learned CLF $\hat{V}(\cdot, \cdot)$ approximates $V^*(\cdot)$ well in areas of the state space, where training data is provided. Moreover, the induced closed-loop system is rendered asymptotically stable. Given a dense-enough sampling, the stabilizing policy $\bm{\hat{\pi}}(\cdot, \cdot)$ induces similar dynamics as the optimal policy $\bm{\pi}^*(\cdot)$. Note that the first column illustrates the initial guess $\hat{V}_0(\cdot, \cdot)$, which only fulfills \eqref{eq:prop:gradsquare} and does not need to constitute a valid CLF. }
    \label{fig:ground_truth}
\end{figure*}

\subsection{Ground Truth Evaluation}

\Cref{fig:ground_truth} shows the obtained results. In the top row of \Cref{fig:ground_truth}, the contours of the ground truth value function $V^*(\cdot)$ are shown in blue (dashed) and the contours of the learned CLF $\hat{V}(\cdot, \cdot)$ are shown in green. The grid of observed one-step trajectories is visualized using purple stars, where the star indicates the starting point $\bm{x_i}$ of each observation tuple. % $\big\{\bm{x_i}, \bm{f}^*(\bm{x_i}) + \bm{\sigma}(\bm{x_i}) \big\}$. 
Conversely, data free areas are illustrated by the greyed out parts of the state space. The bottom row of \Cref{fig:ground_truth} depicts the resulting closed-loop dynamics induced by the stabilizing control law $\bm{\hat{\pi}}(\cdot, \cdot)$ \eqref{eq:final_sontag} modulated by the gradient of the respective CLF $\hat{V}(\cdot, \cdot)$. From left to right, each column of \Cref{fig:ground_truth} shows the obtained results for grids spanning a progressively larger area of the state space. 
    \begin{description} [style=unboxed,leftmargin=0cm]
        \setlength\itemsep{0em}
        \item[]The first column depicts the datafree case with the initial guess $\hat{V}_0(\cdot, \cdot)$. As described in \Cref{lem:gs}, $\hat{V}_0(\cdot, \cdot)$ only fulfills condition \eqref{eq:prop:gradsquare} and does not need to constitute a valid CLF. Thus, the related vector field only differ slightly from the open-loop system dynamics and remain unstable.
        
        \item[]In the second column a $7 \times 7$ grid spanning equidistantly over the lower left quadrant of the normalized state space, i.e., $\bm{x} \in [-1, 0]^2$, is used to generate training data. Here, it can be seen that the learned $\hat{V}(\cdot, \cdot)$ agrees with the value function $V^*(\cdot)$ in the shape of their level curves in areas of the state space, where training data is provided. Moreover, the induced closed-loop dynamics using the stabilizing control law $\bm{\hat{\pi}}(\cdot, \cdot)$ indicate a good reproduction of the dynamics under the optimal policy $\bm{\pi}^*(\cdot)$ for the lower left quadrant. On the other hand, in the data free area the learned $\hat{V}(\cdot, \cdot)$ still constitutes a CLF and induces a control policy that globally asymptotically stabilizes the system.
        
        \item[]In column three the dataset is extended to a $7 \times 14$ grid resulting in 98 samples covering the left half of the normalized state space. With increasing sampling density a better estimate of $V^*(\cdot)$ is obtained, resulting in an improved reconstruction of the observed behavior over a larger area. 
        
        \item[] Finally, column four shows a grid sampling over the complete task space. Our proposed approach yields an estimate $\hat{V}(\cdot, \cdot)$ that matches the ground truth value function $V^*(\cdot)$ in the shape of the contour lines up to some small deviations over the whole work space. The system is stabilized under the policy induced by $\hat{V}(\cdot, \cdot)$ and the closed-loop dynamics resemble the ones generated by the optimal policy $\bm{\pi}^*(\cdot)$. \looseness=-1
    \end{description}

Thus, it can be seen that our method is able to retrieve the shape of the true optimal value function $V^*(\cdot)$, if provided with sufficient training data. Additionally, the learned CLF $\hat{V}(\cdot, \cdot)$ is directly accompanied by a control law $\hat{\bm{\pi}}(\cdot, \cdot)$, which generates dynamics that closely resembles the demonstration under the optimal policy $\bm{\pi}^*(\cdot)$. Intuitively, the CLF $\hat{V}(\cdot, \cdot)$ is optimized to modulate dynamics with an attractor landscape matching the observed demonstrations. Since the control law $\hat{\bm{\pi}}(\cdot, \cdot)$ is asymptotically stabilizing by design, regularity properties are retained by generalizing to unseen areas of the state space by guaranteeing robust convergence to the desired goal state. Moreover, from column two and three in \Cref{fig:ground_truth} it can be seen that the learned CLF $\hat{V}(\cdot, \cdot)$ does not agree with the value function $V^*(\cdot)$ in the data free area, which indicates that the achieved similarity in data rich areas is not artificially enforced due to the parametric structure of $\hat{V}(\cdot, \cdot)$, thus, demonstrating the expressive capabilities of the proposed approach. \looseness=-1

\section{Learning from Human Demonstrations}\label{sec6}
While the previous section demonstrates the principle capacity of the learned CLF to approximate an optimal value function, this subsection shows the expressiveness of our method by learning from human demonstrations in a goal-directed movement task. % and comparing the achieved performance to a state-of-the-art IRL method. 
To this end, we perform a benchmark evaluation in which we compare our approach to one state of the art IRL and DMP method using a human handwriting data set \cite{Khansari11}.

\subsection{Comparison Methods}
Since our proposed approach conceptually unifies the advantages of IRL and DMP methods, we include one related work of each type here for the performed comparison evaluation. Specifically, we consider adversarial inverse reinforcement learning (AIRL) \cite{Fu17}, which is an efficient IRL algorithm applicable in continuous state-action spaces and capable of recovering reward functions that are robust with respect to environment dynamics. % that recovers a reward function from a set of state-action demonstrations. 
Here, both the reward function as well as the policy are parameterized by deep neural networks. Since AIRL is a sampling-based method, %it requires evaluation of the current policy estimate in every iteration, however, 
it does not require an analytic description of the open-loop dynamics. In contrast to our approach, AIRL infers a reward function, i.e, the stage-costs, which generally contains more information than the value function but does not provide any inherent stability guarantees for the learned policy. %Moreover, state-action-pairs are needed for the training process in AIRL.
Secondly, we compare our approach to a DMP method called CLF-DM (Control Lyapunov Function-based Dynamic Movements) \cite{Khansari2014}, where a task-oriented, CLF-like function is learned from data to generate stable dynamical systems using a virtual, stabilizing controller.  
%Similarly to our approach, a task-oriented CLF is learned from demonstrations to guarantee the global asymptotic stability of the resulting dynamical system. 
The process of learning the CLF, the dynamics and the virtual controller is split into three distinct steps, thereby the motion encoding and stabilization are separated, which is in contrast to our integrated approach. Moreover, \cite{Khansari2014} is a DMP method, and thus, only generates motion plans and not an actual control policy that can be applied to a real system. \looseness=-1

\subsection{Demonstration Data and Evaluation Setup}
\begin{figure}
        \centering
        \includegraphics[width=.425\textwidth]{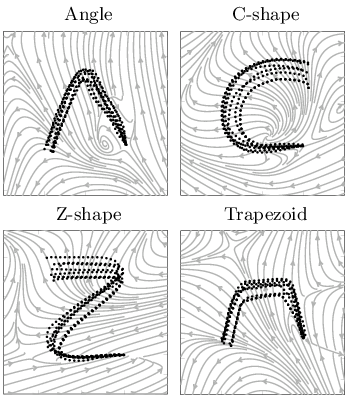}
       \caption{Depiction of 4 exemplary shapes of the LASA handwriting dataset. %and the associated open-loop dynamics. Here, the uncontrolled system, illustrated by gray streamlines, is unstable and governed by a polynomial system of degree 3. 
       The associated, unstable open-loop dynamics (gray) are polynomial functions of degree 3. The demonstrations shown with black dots are generated by an agent acting on the dynamical system. \looseness=-1}
      \label{fig:LASA_open_loop}
    \vspace{-0.3cm}
    \end{figure}

We evaluate all methods on the real-world LASA handwriting data set \cite{Khansari11}, which consists of 30 human-drawn trajectories of various letters and shapes with 7 demonstrations each. The data set is commonly used to compare approaches for learning from demonstration. All data is normalized to a state space $\pazocal{X} \in [-1, 1]^2$. %We use the code provided by the authors of the respective algorithms. 
%Since both AIRL and our approach are able to learn a control policy for a given system, 
While the LASA data set itself does not provide information regarding the dynamics, we consider unstable, polynomial systems of degree 3 for the open-loop environment dynamics of each shape. Thereby, it is emulated that the provided demonstrations represent an agent acting on a dynamical system and allows for a non-trivial modelling of the agent by means of the inferred value function. %To this end, AIRL is chosen as a comparison method, as the learned reward function is robust to influence of environment dynamics and the inference is efficient and expressive \cite{Fu17}. 
The resulting open-loop dynamics and the demonstration data are illustrated for 4 exemplary shapes in \Cref{fig:LASA_open_loop}.
%For the training of the AIRL algorithm $T=1000$ samples per demonstration are used, while our algorithm is trained on $T=100$ samples per demonstration, as it did not show a meaningful degradation in tracking performance on a reduced data set. In both cases, the number of observed trajectories is $N=7$ per shape, leading to a total data set size of $7000$ per shape for the AIRL method and $700$ for our approach. 
For the training $N=7$ demonstrations are observed per shape, where each demonstration trajectory includes $T=1000$ samples resulting in a total data set size of $7000$ per shape.
Due to the stochastic initialization strategies of the AIRL algorithm, training is performed for 5 different seeds with $1000$ iterations each, from which the seed with the best tracking performance is selected for the comparison evaluation. The network size is chosen to 3 hidden layers, each consisting of $64$ neurons for both the reward and the policy network. For CLF-DM we use the proposed Weighted Sum of Asymmetric Quadratic Function (WSAQF) with $\mathcal{L}=3$ asymmetric functions to learn the energy function. Furthermore, a Gaussian Mixture Regression consisting of 8 components is used to fit the dynamical system. Finally, for our proposed approach, a degree of $\text{deg}(\hat{V})=8$ is chosed for $\hat{V}(\bm{c},\cdot)$ and $\text{deg}(\lambda)=16$ for $\lambda(\bm{\theta},\cdot)$. Algorithm 1 is initialized with $\hat{V}_0(\bm{x})=x_1^2 + x_2^2$ and $\bm{\lambda}_0(\bm{x}) = 0.5 \bm{I}_2$ and configured to run for 5 iterations. Moreover, the regularization factor is set to of ${\bm{\Sigma}_c^{-1} = 10^{-4}\bm{I}_2}$. % for $\gamma_V$ and $\gamma_\lambda$.
 %The resulting closed loop dynamics for each algorithm are simulated with the respective discrete integration time provided for each shape in the LASA data set.
The reproduction performance is evaluated using three similarity measures, i.e., mean squared error (MSE), piecewise curve mapping (PCM) and dynamic time warping (DTW), for which the implementation in \cite{Jekel19} is used. \looseness=-1 %The mean and standard deviation for each similarity measure is calculated for all 7 reproductions and shown in figure \ref{fig:eval:COMP:measures}. 

\newcommand{\specialcell}[2][c]{%
  \begin{tabular}[#1]{@{}c@{}}#2\end{tabular}}

\begin{figure*}
    \centering
    \includegraphics[width=.9\textwidth]{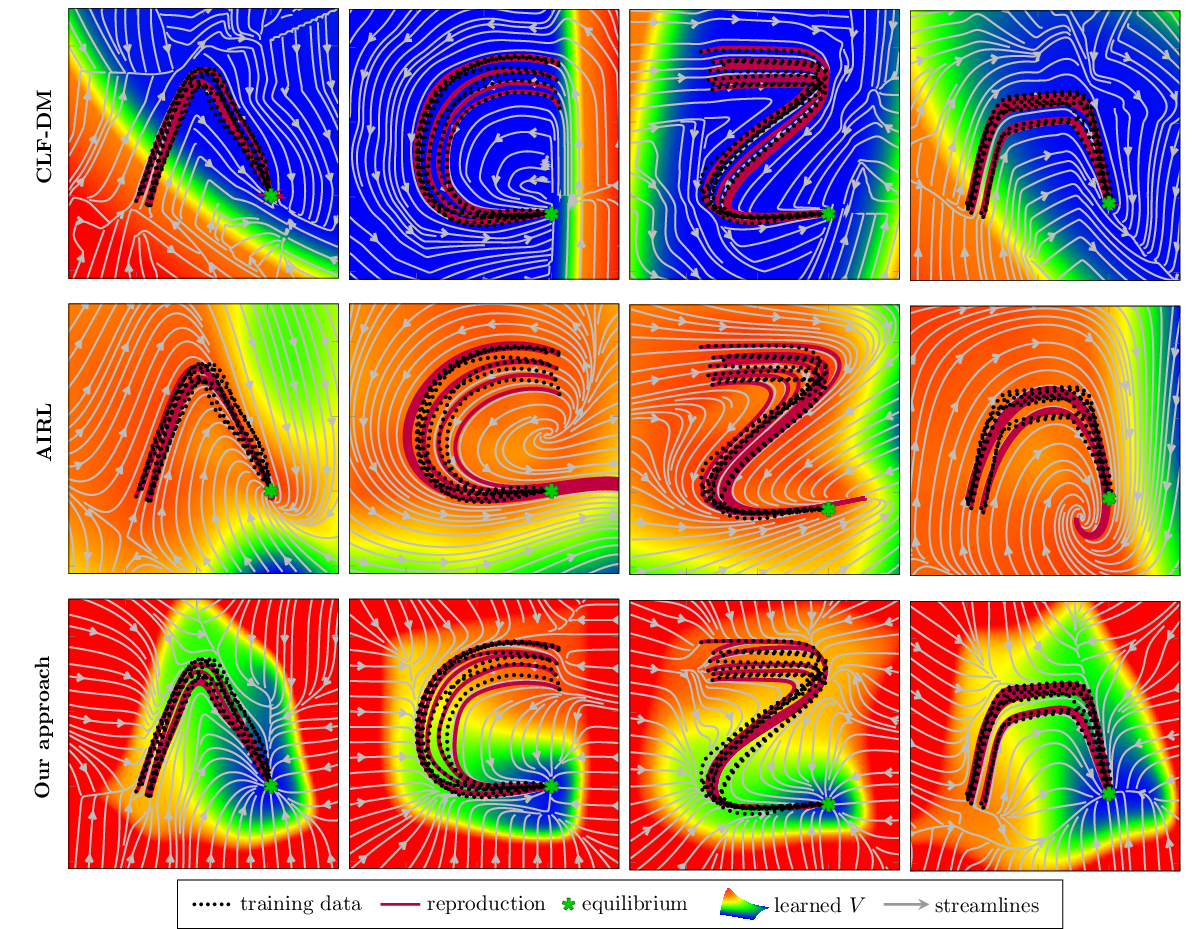}%{pdfs/placeholder/Comparison_full.JPG}
    \caption{Comparison of the learned $\hat{V}(\cdot,\cdot)$ and reproductions for the our proposed approach, AIRL and CLF-DM. The reproductions are illustrated by a red line and the training data is represented by black dots. The learned potential functions are visualized in the form of a projected surface plot underlying the streamlines (gray arrows). Here, the shape encoding in the learned CLF is clearly visible for our approach, as the potential function shrinks along the vector field and the minimum coincides with the target state. Moreover, our method demonstrates convergence to the equilibrium for all shapes. \looseness=-1}
    \label{fig:comparison}
\end{figure*}

\subsection{Benchmark Evaluation}
The results of the comparison evaluation are depicted for the 4 exemplary shapes in \Cref{fig:comparison}. Here, both the learned value function, in the case of AIRL, and learned Lyapunov function for CLF-DM and our approach are visualized together with the resulting vector field. % of the closed-loop dynamics under the associated policy. 
Note that the vector fields for our method and AIRL indicate the closed-loop dynamics under the associated policy, while the ones shown for CLF-DM merely represent the learned, virtual dynamics and not a controlled, closed-loop system.

\begin{description} [style=unboxed,leftmargin=0cm]
\item[ Reproduction performance ] 
When inspecting the reproduced trajectories, it can be seen directly that our method and CLF-DM induce asymptotic convergence to the desired equilibrium for all shapes. In contrast, AIRL produces unstable closed-loop dynamics for the C-shape %and Sine or 
and convergence to a spurious attractor in the case of the Z-shape and Trapezoid. As there are no guarantees for the learned policy, this behavior varies for each scenario. Hence, even in cases where the reproductions converge to the desired equilibrium, no statement regarding the region of attraction or generalization properties can be made. 
%\textcolor{red}{Moreover, CLF-DM also produces asymptotic convergence to the equilibrium only for some shapes, while others appear to diverge. Here, it is unclear if the generated trajectories would converge given a longer simulation time.} 
When inspecting the reproductions it can be seen that %the AIRL method seems to encode complex movements more precisely in some cases due to the expressiveness of the neural network function approximator. On the other hand, 
CLF-DM and our approach generate more dynamic motion, while AIRL generally induces smoother trajectories. However, in particular for highly dynamic demonstrations, our method exhibits a \textit{skipping} behavior in some instances, where a conservative convergence to the equilibrium is induced. %, e.g., for Multi-modal 4 and Sine. %This can be explained by a limited expressiveness due to the degree of the polynomial function approximating the CLF. 
\end{description}

\begin{figure}
    \centering
    \includegraphics[width=.485\textwidth, trim={9mm 0 0 0}, clip]{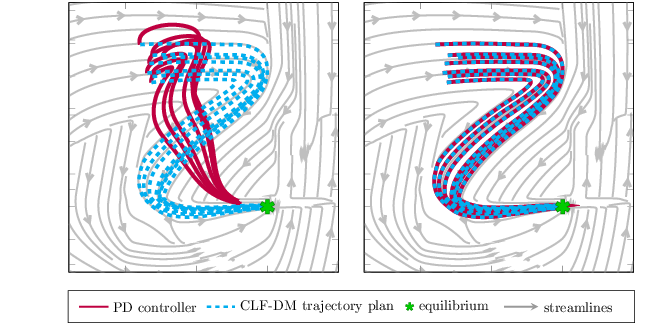}
    \vspace{-0.25cm}
    \caption{Comparison of two control parametrization, i.e., (left) a low-gain and (right) a high-gain PD controller, to track the trajectory plan generated by CLF-DM for the Z-shape. It can be seen that the reproduction performance is highly dependent on the design choice of the tracking controller. }
    \vspace{-0.3cm}
    \label{fig:pd_tracking}
\end{figure}

\begin{description} [style=unboxed,leftmargin=0cm]
\item[ Quantitative evaluation ]
In order to perform a more rigorous comparison, a quantitative evaluation is provided in \Cref{table:comparison_results}. Here, we compare the reproduction similarity %in terms of mean squared error (MSE), partial curve matching (PCM) and dynamic time warping (DTW) 
for the respective shapes depicted in \Cref{fig:comparison} together with the average over all 30 shapes included in the dataset. %In \Cref{table:comparison_results} i
It can be seen that AIRL exhibits a better performance according to the MSE. However, for similarity measures that are strongly related to the geometric accuracy of curves (PCM and DTW), our method consistently outperforms AIRL. Moreover, due to a lack of convergence guarantees, the reproduction quality of AIRL suffers greatly if task success is explicitly linked to reaching a specific goal state exactly, which cannot be expressed by the similarity measures depicted in \Cref{table:comparison_results}. Therefore, our approach generates superior reproductions in terms of geometric similarity to the observed demonstrations and guaranteed convergence to the desired target state. 
Note that the measures achieved by CLF-DM are grayed out in \Cref{table:comparison_results}, since it merely represents a trajectory plan and the achieved reproduction primarily depends on the design of a subsequent tracking controller. This can be seen in \Cref{fig:pd_tracking} which illustratively depicts the achieved tracking performance when deploying a PD controller $\bm{u}_{\text{PD}}(\bm{x},\dot{\bm{x}})\coloneqq\bm{K}_P(\bm{x}_\text{ref}-\bm{x}) + \bm{K}_D(\bm{\dot{x}}_\text{ref}-\bm{\dot{x}})$ to track the trajectory plan $\{\bm{x}_\text{ref}, \bm{\dot{x}}_\text{ref}\}$ generated by CLF-DM for two different control gain parametrization, i.e., a low-gain parametrization $\bm{K}_{P,\text{low}}=0.5\bm{I}_2$, $\bm{K}_{D,\text{low}}=0.05\bm{I}_2$
% \begin{equation}
% 	\bm{K}_{P,\text{low}} = 
% 	\begin{bmatrix}
% 	0.5     &  0 \\
% 	0       &  0.5
% 	\end{bmatrix},
% 	\enspace
% 	\bm{K}_{D,\text{low}} = 
% 	\begin{bmatrix}
% 	0.05    &  0 \\
% 	0       &  0.05
% 	\end{bmatrix}, \nonumber
% \end{equation}
and a high-gain parametrization $\bm{K}_{P,\text{high}}=50\bm{I}_2$, $\bm{K}_{D,\text{high}}=1.0\bm{I}_2$.
% \begin{equation}
% 	\bm{K}_{P,\text{high}} = 
% 	\begin{bmatrix}
% 	50     &  0 \\
% 	0      &  50
% 	\end{bmatrix},
% 	\enspace
% 	\bm{K}_{D,\text{high}} = 
% 	\begin{bmatrix}
% 	1    &  0 \\
% 	0    &  1
% 	\end{bmatrix}. \nonumber
% \end{equation}
It is clearly visible that the achieved reproductions primarily depend on the design of the tracking controller which is not trivial in general. Thus, a direct comparison of the output provided by CLF-DM and the one generated by our approach and AIRL with respect to reproductions is difficult.
\end{description}

\newcolumntype{L}[1]{>{\raggedright\let\newline\\\arraybackslash\hspace{0pt}}m{#1}}
\newcolumntype{C}[1]{>{\centering\let\newline\\\arraybackslash\hspace{0pt}}m{#1}}
\newcolumntype{R}[1]{>{\raggedleft\let\newline\\\arraybackslash\hspace{0pt}}m{#1}}
\csvstyle{mystyle}{
    % table head=\toprule,
    % table foot=\bottomrule,
    % no head,
    late after line=\\,
    % late after first line=\\\midrule,
    }
% Width for each picture
\newcommand\ecs{0.085\textwidth}

\begin{description} [style=unboxed,leftmargin=0cm]
\item[ Intention encoding ] 
Besides obtaining a policy that replicates the observed demonstrations, one of the key benefits of IRL methods is the inference of a cost function representation, which facilitates interpretability by modelling the underlying agent intention.
%In order to obtain interpretable human motion models, it is beneficial to retrieve an accurate representation of the functions driving the predictions, i.e., the underlying intention. 
In \Cref{fig:comparison} the associated potential functions, e.g., value function, task-oriented CLF or CLF-based value function approximation, represent this intention. 
For the CLF-DM method the shapes are not attributable to the task-oriented CLF, even though it is learned from the demonstrations. %Here, the motion encoding is performed by a, potentially unstable, dynamic system in combination with a virtual stabilizing control \myref{[KZ14]} \textcolor{red}{[Khansari11]?}. Thus, 
Since the process of learning the CLF, the dynamic system and the virtual, stabilizing controller is separated, the coherence of the retained CLF to a given reproduction result is not guaranteed.   %retrieve a cost function, which constitutes a rich representation of the intention, the
On the other hand, despite the cost function inference performed by AIRL, the resulting value functions depicted in \Cref{fig:comparison} is not unambiguously relatable to the demonstrated shapes. This is since nonlinear IRL methods, such as \cite{FinnGCL16, FinnGAN16, Fu17}, only approximately solve the computationally demanding forward optimal control problem that iteratively occurs during the cost function inference. In the case of AIRL for instance, only a few RL iterations are performed to learn the optimal policy for the current reward function parametrization instead of running the algorithm to convergence, thereby accelerating the inference process. However, while the pair of retrieved cost function and policy reproduce the demonstrations well, the two functions are no longer consistently related due to the performed approximation. Hence, the inferred cost function may not represent the underlying intention anymore. 
In contrast, for our approach the demonstrated shapes are clearly visible in the projected surface plots of the learned CLF-based value function approximation. This is since our method completely encodes the generated motion through the Lyapunov function, as the closed-form stabilizing policy $\bm{\hat{\pi}}(\cdot,\cdot)$ is directly dependent on the gradient of the CLF. By exploiting this analytical link it is possible to directly search over the space of parametrization of $\hat{V}(\cdot,\cdot)$ by optimizing the attractor landscape of the generated closed-loop dynamical system to capture the observed demonstration. Thereby we achieve an efficient and fully integrated learning process, where the inferred CLF induces a control vector field that reproduces the demonstrations, while also guaranteeing asymptotic convergence to the task goal by design. Moreover, through the closed-form expressions we obtain consistently interpretable solutions. 
\end{description}

\begin{table*}[htbp]
\footnotesize
 \centering
  \begin{threeparttable}
        \caption{Resulting mean squared error (MSE), dynamic time warping (DTW) and piecewise curve mapping (PCM) measures over 4 shapes and the average over all shapes. For each measure, the mean along with the standard deviation calculated from all 7 reproductions per shape is given. %\textcolor{red}{The calculation of the PCM and DTW measure for the AIRL algorithm on the Sine-shape was reduced to 1500 samples per reproduction due to numerical difficulties calculating the curve mapping for the unsable reproduction.}
        }
        \centering
        \begin{tabular}{L{0.076\textwidth} C{0.0525\textwidth} C{0.05\textwidth} C{0.0675\textwidth} C{0.08\textwidth} C{0.095\textwidth} C{0.0775\textwidth} C{0.08\textwidth} C{0.095\textwidth} C{0.08\textwidth}}
            % Header: Algorithm names
            \toprule
             & \multicolumn{3}{c}{MSE $\times10^3$} & \multicolumn{3}{c}{DTW} & \multicolumn{3}{c}{PCM}  \\ %& \multicolumn{2}{c}{Our Approach}
            % Spacer: Some horizontal lines indicating grouping
            \cmidrule(lr){2-4}\cmidrule(lr){5-7}\cmidrule(lr){8-10}
            & Our & AIRL & CLF-DM & Our & AIRL & CLF-DM & Our & AIRL & CLF-DM \\
            \toprule
            % Standard dataset (1000 samples) & & & & & & \\
            % \midrule
            \begin{filecontents*}{evaluation_data/measures_unextended.csv}
shapename,mse_sosclf_mean,mse_sosclf_std,mse_irl_mean,mse_irl_std,mse_dmp_mean,mse_dmp_std,dtw_sosclf_mean,dtw_sosclf_std,dtw_irl_mean,dtw_irl_std,dtw_dmp_mean,dtw_dmp_std,pcm_sosclf_mean,pcm_sosclf_std,pcm_irl_mean,pcm_irl_std,pcm_dmp_mean,pcm_dmp_std
Angle,11,9\phantom{0},\textbf{3},\textbf{2},12,13,\textbf{35.7},\textbf{17.9},68.6,13.4,36.4,19.3,29,15,65.4,7.2\phantom{0},\textbf{27.9},\textbf{16\phantom{.0}}
CShape,31,21,\textbf{7},\textbf{5},21,13,38.7,22.2,689.3,27.0\phantom{0},\textbf{29.9},\textbf{13\phantom{.0}},26.3,14.5,662.3,35.9\phantom{0},\textbf{21.5},\textbf{8.8\phantom{0}}
Zshape,30,23,\textbf{11},\textbf{7}\phantom{0},15,16,\textbf{46.9},\textbf{19.5},292.9,32.6\phantom{0},57.3,28\phantom{.0},\textbf{38.3},\textbf{23.7},277.2,81.5\phantom{0},47.1,23.9
Trapezoid,4,3,\textbf{2},\textbf{1},3,3,\textbf{22.5},\textbf{10.2},241.9,6.2\phantom{00},32.6,8.4\phantom{0},\textbf{19.2},\textbf{10.5},344.1,17.2\phantom{0},26.7,10.3
\end{filecontents*}
\csvreader[mystyle]{evaluation_data/measures_unextended.csv}{}% use head of csv as column names
            {\csvcoli & \csvcolii${\pm}$\csvcoliii & \csvcoliv${\pm}$\csvcolv & \textcolor{gray}{\csvcolvi${\pm}$\csvcolvii} & \csvcolviii${\pm}$\csvcolix & \csvcolx${\pm}$\csvcolxi  & \textcolor{gray}{\csvcolxii${\pm}$\csvcolxiii} & \csvcolxiv${\pm}$\csvcolxv & \csvcolxvi${\pm}$\csvcolxvii & \textcolor{gray}{\csvcolxviii${\pm}$\csvcolxix}}% specify your coloumns here
            
            % Angle & $11{\pm}9\phantom{0}$ & $3{\pm}2$ & $12{\pm}13$ & $35.7{\pm}17.9$ & $68.6{\pm}13.4$ & $36.4{\pm}19.3$ & $29.0{\pm}15.0$ & $65.4{\pm}7.2$ & $27.9{\pm}16.0$ \\ 
            
            % CShape & $31{\pm}21$ & $7{\pm}5$ & $21{\pm}13$ & $38.7{\pm}22.2$ & $689.3{{\pm}}27.0$ & $29.9{\pm}13.0$ & $26.3{\pm}14.5$ & $662.3{\pm}35.9$ & $21.5{\pm}8.8$ \\ 

            % ZShape & $30{\pm}23$ & $11{\pm}7\phantom{0}$ & $15{\pm}16$ & $46.9{\pm}19.5$ & $292.9{{\pm}}32.6$ & $57.3{\pm}28.0$ & $38.3{\pm}23.7$ & $277.2{\pm}81.5$ & $47.1{\pm}23.9$ \\   

            % Trapezoid & $4{\pm}3$ & $2{\pm}1$ & $3{\pm}3$ & $22.5{\pm}10.2$ & $241.9{{\pm}}6.2$ & $32.6{\pm}8.4$ & $19.2{\pm}10.5$ & $344.1{\pm}17.2$ & $26.7{\pm}10.3$ \\  
            
            \midrule
            All shapes average & $67{\pm}33$ & $\textbf{8}{\pm}\textbf{5}$ & $13{\pm}12$ & $124{\pm}66.4$ & $333{\pm}33.9$ & $\textcolor{gray}{\textbf{70}{\pm}\textbf{31.5}}$ & $85.5{\pm}35.5$ & $334.3{\pm}41.5$ & $\textcolor{gray}{\textbf{54.5}{\pm}\textbf{30.5}}$ \\
            \bottomrule
        \end{tabular}
        \label{table:comparison_results}
    \end{threeparttable}
\end{table*}

\section{Conclusion} \label{sec6}
In this article, we propose a novel approach for performing stable IRL by exploiting the inverse optimal relation between the value function and control Lyapunov functions. In particular, we reformulate the cost function inference to learning a Lyapunov-constrained value function approximation. For the continuous evaluation of the value function, we propose the use of gradient-based, stabilizing control laws leading to a closed-form expression of the log-likelihood. Due to the analytic expressions, the optimization of $\hat{V}(\cdot,\cdot)$ is intuitively analogue to a search over the space of point-attractor landscapes spanned by the closed-loop dynamics. We perform a theoretical analysis for the CLF-based approximation and demonstrate under which conditions optimality properties are retained. 
Additionally, we present how inverse optimal CLFs are efficiently learned from data using the SOS technique. First, %in \Cref{lem:err_schur}, 
an equivalent, convex reformulation of the posterior expressions is derived. Second, the Lyapunov-constrained problem is convexified by proposing a sequential, two-step optimization approach. 
The algorithm allows for the efficient inference of CLFs $\hat{V}(\cdot,\cdot)$ and guarantees that the closed-loop dynamics under policy $\hat{\bm{\pi}}(\cdot,\cdot)$ are asymptotically stable. In the evaluation, we show that our proposed framework exhibits the desired convergence properties and performs superior in encoding the preferences of a human demonstrator %based on a real-world data set 
in comparison to a state-of-the-art IRL and DMP method.

\bibliographystyle{IEEEtran}
\bibliography{references}

\end{document}